\documentclass[12pt]{iopart}

\usepackage{graphics}
\newcommand{\auno}{{\rm A}\ensuremath{_1}}

\newcommand{\buno}{{\rm B}\ensuremath{_1}}
\newcommand{\bdue}{{\rm B}\ensuremath{_2}}
\newcommand{\ebid}{{\rm E}}

\newcommand{\cebid}{\ensuremath{{\mathcal E}}}

\newcommand{\Bcr}[1]{{\it b}\ensuremath{^{\dagger}_{#1}}}

\newcommand{\Ccr}[1]{{\it c}\ensuremath{^{\dagger}_{#1}}}

\newcommand{\Dcr}[1]{{\it d}\ensuremath{^{\dagger}_{#1}}}

\newcommand{\Pcr}[1]{{\it p}\ensuremath{^{\dagger}_{#1}}}

\newcommand{\be}{\begin{equation}}
\newcommand{\ee}{\end{equation}}
\newcommand{\beq}{\begin{eqnarray}}
\newcommand{\eeq}{\end{eqnarray}}
\newcommand{\cu}[0]{\mathrm{CuO}_{4}}
\newcommand{\cuoq}[0]{\mathrm{CuO}_{4} }
\newcommand{\ve}{\Vertex}

\begin{document}

\def\gC{\mbox{\boldmath $C$}}
\def\gZ{\mbox{\boldmath $Z$}}
\def\gR{\mbox{\boldmath $R$}}
\def\gN{\mbox{\boldmath $N$}}
\def\cN{{\cal N}}
\def\ua{\uparrow}
\def\da{\downarrow}
\def\eq{\equiv}
\def\a{\alpha}
\def\b{\beta}
\def\g{\gamma}
\def\G{\Gamma}
\def\d{\delta}
\def\D{\Delta}
\def\e{\epsilon}
\def\ve{\varepsilon}
\def\z{\zeta}
\def\h{\eta}
\def\th{\theta}
\def\k{\kappa}
\def\l{\lambda}
\def\L{\Lambda}
\def\m{\mu}
\def\n{\nu}
\def\x{\xi}
\def\X{\Xi}
\def\p{\pi}
\def\P{\Pi}
\def\r{\rho}
\def\s{\sigma}
\def\S{\Sigma}
\def\t{\tau}
\def\f{\phi}
\def\vf{\varphi}
\def\F{\Phi}
\def\w{\omega}
\def\W{\Omega}
\def\Q{\Psi}
\def\q{\psi}
\def\de{\partial}
\def\inf{\infty}
\def\ra{\rightarrow}
\def\bra{\langle}
\def\ket{\rangle}

% Journal identifier can be put here if required, e.g.
\jl{3}

\topical[$W=0$ pairing in Hubbard Models]{$W=0$ pairing in Hubbard and
related models of low-dimensional superconductors}

\author{Adalberto Balzarotti *, Michele Cini *\dag\footnote[1]{To
whom correspondence should be addressed.} and Enrico Perfetto
*\dag, and Gianluca Stefanucci\ddag\ }

\address{
*\ Istituto Nazionale per la Fisica della Materia, Dipartimento di
Fisica,
 Universita' di Rome Tor
Vergata, via della ricerca Scientifica 1, 00133 Roma Italy \\
\dag\ INFN - Laboratori Nazionali di Frascati, Via E. Fermi 40,
00044 Frascati, Italy}

\address{\ddag\ Department of Solid State Theory, Institute of
Physics, Lund University, S\"olvegatan 14 A, 22362 Lund Sweden}

\begin{abstract}
Lattice Hamiltonians with on-site interaction $W$ have
$W=0$ solutions, that is, many-body {\em singlet} eigenstates without double
occupation. In particular, $W=0$ pairs give a clue to understand the
pairing force in repulsive Hubbard models.
These eigenstates are found in systems with
high enough symmetry, like the square, hexagonal or triangular
lattices. By a general theorem, we propose a systematic way to
construct all the $W=0$ pairs of a given Hamiltonian.
We also introduce a canonical transformation to calculate
the effective interaction between the particles of such pairs.
In geometries appropriate for the CuO$_{2}$ planes
of cuprate superconductors, armchair Carbon nanotubes or Cobalt
Oxides planes, the dressed pair becomes a bound state in a physically
relevant range of parameters. We also show that $W=0$ pairs quantize the
magnetic flux like superconducting pairs do. The pairing mechanism breaks
down in the presence of strong distortions.
The $W=0$ pairs are also the building blocks for the antiferromagnetic ground
state of the half-filled Hubbard model at weak coupling.
Our analytical results for the $4\times 4$ Hubbard square lattice,
compared to available numerical data,
demonstrate that the method, besides providing intuitive grasp on
pairing, also has quantitative predictive power.
We also consider including phonon effects in this scenario.
Preliminary calculations with  small clusters indicate  that vector phonons
hinder pairing  while half-breathing modes are
synergic with the $W=0$ pairing mechanism both at weak coupling
and in the polaronic regime.
\end{abstract}


\tableofcontents

\section{Introduction}
\label{section1}

In the last decades, much effort has been devoted to  {\em exotic}
mechanisms of pairing and to the  possibility of non-conventional
superconductivity in new materials. In a  list of the most
important  and frequently discussed examples, one could mention
the cuprates\cite{bedmu}, organic superconductors including
fullerenes\cite{fullerene} and carbon nanotubes\cite{tang},
ruthenates\cite{ruthe}, and Na-Co oxides\cite{takada}. Among the
most interesting possibilities, correlation effects have been
invoked, and the Hubbard model\cite{hubbard} has been increasingly
popular to achieve a simplified picture.

The Hubbard model is believed to
exhibit various interesting phenomena including antiferromagnetism,
ferrimagnetism, ferromagnetism, metal-insulator transitions, etc.
It is the simplest  Hamiltonian covering both aspects of a strongly correlated
electron system (like the CuO$_{2}$ planes of cuprates),
namely the competition between band-like behaviour and the tendency to
atomic-like localisation driven by the screened Coulomb repulsion.
Besides, several authors believe that it  can exhibit a superconducting
phase in a certain parameters regime. Despite its simplicity, the Hubbard
Hamiltonian cannot be exactly solved in more than 1d and a large variety
of approaches has been proposed to study the superconducting correlations
in the ground state and at finite temperature.
Bickers and co-workers\cite{bickers2}
were among the first to propose the $d_{x^{2}-y^{2}}$ wave
superconductivity in 2d.
A recent survey of the superconducting properties
of the single- and multi-band Hubbard model can be found in Ref.\cite{hotta}.

Another popular class of models are that obtained from the Hubbard
Hamiltonian in the strong coupling limit. A large on-site Coulomb
repulsive energy $U$ between carriers of opposite spins tends to
reduce the double occupancy. A straightforward perturbation theory
in the parameters $t$ of the kinetic term leads to Heisenberg-like
Hamiltonians for fermions propagating through an antiferromagnetic
background with an exchange interaction $J$ between neighbouring
spins. These are the so called t-J
models\cite{dagrev},\cite{sush}. In the t-J model the
double-occupancy is forbidden by the so called Gutzwiller
projection\cite{gutz}, and there is no on-site repulsion. However,
superconductivity is a delicate phenomenon and the exclusion of
double occupied sites costs kinetic energy, so the t-J model might
fail to  give an appropriate description of the Hubbard model when
$U$ is comparable to $t$.

In this review article, we illustrate the ``$W=0$'' pairing
mechanism, in which  symmetry is capable of cutting down the
on-site repulsion from the outset, without any need of the
Gutzwiller projection. This effect works at any $U/t$ and it
relies on a configuration mixing which entails the presence of
degenerate one-body states at the Fermi level. Moreover, we
provide evidence that this configuration mixing, when applied to
the full many-body problem, can produce pairing when physical
parameter values are used.

The pairing force in the Hubbard model is induced  by repulsive
interactions, and recalls the Kohn-Luttinger\cite{kohn} pairing in
the Jellium. They pointed out that any three-dimensional Fermi
liquid undergoes a superconducting instability by Cooper pairs of
parallel spins and very large relative angular momentum $l$. A
simplified view of the Kohn-Luttinger effect is given by
considering one particle of the pair as an external charge. Then,
the screening gives rise to a long-range oscillatory potential
(Friedel oscillations) due to the singularity of the longitudinal
dielectric function at 2$k_{\rm F}$; here $k_{\rm F}$ is the Fermi
wavevector. The strict reasoning exploits the fact that the
Legendre expansion coefficients of any regular direct interaction
between particles of opposite momentum drops off exponentially in
$l$. On the other hand, the second-order contribution to the
scattering amplitude falls as $1/l^{4}$ and at least for odd $l$
leads to an attractive interaction. In the modern renormalization
group language\cite{shankar}, the second-order correction is
obtained by summing up the marginal scattering amplitudes of the
isotropic Fermi liquid coming from the so-called Forward channels,
including, for antiparallel spins, a spin-flip diagram. This
scenario does not work in the two-dimensional Fermi liquid, but
going beyond the second-order perturbation theory the
Kohn-Luttinger effect is recovered\cite{chubukov}.

The present mechanism works with singlet pairs and differs from
Kohn-Luttinger one in other  important ways. In contrast with the
homogeneous electron gas, the lattice structure gives very tight
pairs and the $W=0$ mechanism displays very clearly in tiny
clusters as well. On top of that, it is worth noticing that in
high-T$_{C}$ superconductors the size of the Cooper pairs is
expected to be of the order of few lattice constants and hence the
pairing mechanism should lend itself to cluster studies.
Macroscopically large lattices with periodic boundary conditions
have large symmetry groups including the space groups (point
symmetry + translational symmetry); in such conditions, the $W=0$
pairing mechanism is at its best. Similarly, in finite geometries,
the largest binding energy is obtained in fully symmetric clusters
while static distortions tend to unbind the pair. Exact
calculations on finite models should bring to light interesting
aspects of the microscopic origin of the pairing mechanism, and be
useful as tests for the analytic developments.

To test the superconducting nature of the pairs arising from repulsive
interactions, one can use finite systems in {\em gedanken} experiments.
We probe the behaviour of $W=0$ pairs in the presence of a static
magnetic field and we show that they produce diamagnetic
supercurrents that screen the vector potential. As a result the
superconducting flux quantization is observed in various
geometries.

The paper is organised as follows. In Section \ref{section2}, we
review the pairing mechanism in small symmetric clusters. We prove
the existence of two-body singlet eigenstates with vanishing
on-site Hubbard repulsion, that we call $W=0$ pairs.  We have
collected in  Subsection \ref{section2.2} the  somewhat more
technical aspects, which are central  for the mathematical
foundation of the theory, while readers who are only interested in
the phenomenology might skip it.  We prove a general theorem on
the allowed symmetries of such pairs. From the theorem we extract
a practical recipe to build $W=0$ pairs in any symmetric geometry
(finite or macroscopically large). A careful analysis on the
smallest ``allowed'' cluster, the CuO$_{4}$, shows that pairing
can be obtained in a physical parameter range. The underlying
pairing mechanism is investigated using many-body perturbation
theory. In Section \ref{section3} we generalise the theory to
arbitrary large systems. We introduce a non-perturbative canonical
transformation leading to an effective Hamiltonian for the pair.
The method is free from the limitations of perturbation theory;
the relation of the present formalism to Cooper theory from one
side and to cluster results from the other is discussed. Two kinds
of bound states of different symmetries result, and the dependence
of the binding energy on the filling and other parameters is
explored. In Section \ref{section4} we study the Hubbard Model at
half filling. We remove the ground-state degeneracy in first order
perturbation theory by means of a suitable \textit{local}
formalism. We show that the ground state is the spin singlet
projection of a determinantal state exhibiting the
\textit{antiferromagnetic property}: the translation by a lattice
step is equivalent to a spin flip. As an illustration, the
$4\times 4$ square lattice is studied in detail. The half filled
antiferromagnetic ground state is doped with two holes and an
effective interaction between them is derived by means of the
canonical transformation. The analytical results agree well with
the numerical ones and this shows the predictive power of the
approach. In Section \ref{section5} we investigate the $W=0$
pairing mechanism in carbon nanotubes and triangular cobalt
oxides. Section \ref{section6} is aimed to study the
superconducting magnetic response of symmetric Hubbard models with
$W=0$ bound pairs. Section \ref{section7} deals with the inclusion
of the lattice degrees of freedom; we show that  phonons give a
synergic contribution to the purely electronic mechanism and
catastrophic Jahn-Teller distorsions do not occur. Finally, we
draw our conclusions and outlook in Section \ref{section8}.

\section{Pairing in the Hubbard Model}
\label{section2}

In this review, we shall deal with  Hubbard-like models of various
geometries, designed for application to  superconducting strongly
correlated materials. The prototype Hubbard Hamiltonian reads
\begin{equation}
H = K + W= t \sum_{\langle i,j \rangle , \sigma} c^{\dagger}_{j \sigma}
c_{i\sigma}
+U \sum_{i} n_{i \ua}n_{i \da}  \, ,
\label{1bhub}
\end{equation}
where $K$ stands for the kinetic energy while $W$ accounts for the
on-site repulsive interaction. The summation on $\langle i,j \rangle $
runs over sites $i$ and $j$ which are nearest neighbours.

In Eq. (\ref{1bhub}) the interaction term is repulsive and there
is no electron-phonon coupling, so the very existence of pairing
is a paradox. An effective attractive force comes from, e.g., the
exchange of spin fluctuations but it must be stronger than the
direct Hubbard repulsion to give rise a bound pair. Bickers and
co-workers\cite{bickers2} explored the consequences of a spin
density wave instability on such a pairing force within the RPA
approximation. They found a superconducting phase with
pair-wavefunctions of $d_{x^{2}-y^{2}}$ symmetry in the 2d Hubbard
model. The next level of calculations were carried out by using
the FLEX approximation\cite{bickers}. This method treats the
fluctuations in the magnetic, density and pairing channels in a
self-consistent and conserving way. It was found  that the
antiferromagnetic fluctuations lead to a superconducting phase of
$d_{x^{2}-y^{2}}$ symmetry which neighbours the SDW phase,in
accordance  with the previous findings.

The phase diagram becomes less clear close to half filling because of
the numerous infrared divergencies due to the nesting of the
Fermi surface and Van Hove singularities. As a consequence, the results
of any many-body treatment depend on the choice of diagrams to be summed.
Renormalization Group (RG) methods\cite{shankar} are a well controlled alternative approach
to deal with Fermi systems having competing singularities. The
RG has been used  by several
authors\cite{zanchi},\cite{metzner},\cite{honerkamp} to study the
coupling flows at different particle densities.
In agreement with the previous findings, RG
calculations show a $d$-wave superconducting instability away from half filling.
The underlying physical mechanism, namely exchange of spin- or
charge-density fluctuations, is also the same as in the FLEX approach.

Despite such evidence, there is no general agreement on the
existence of a superconducting phase in the Hubbard
model\cite{anderson3},\cite{anderson4},\cite{susu}. In order to
clarify such a controversy it is very useful to have exact data to
rely on. Unfortunately, the one band Hubbard model is exactly
solved only in one spatial dimension\cite{liebwu} and no sign of
superconductivity was found there\cite{1dl}. Very few exact
analytical results are available for the 2 dimensional
case\cite{2dl}. Therefore in 2 dimensions the low-lying states
must be explored by means of exact diagonalizations on finite
clusters. This kind of numerical calculation, complemented by
analytical work and physical insight, may bring to light
interesting local aspects of the microscopic pairing mechanism.

The exact ground state of  several small Hubbard systems have been
numerically found by means of Lanczos and quantum Monte Carlo
techniques. A  pairing criterion that we shall discuss below  was
given by Richardson in the contest of nuclear
physics\cite{richardson}. Defining
\begin{equation}
{\tilde \Delta}(N+2)= E(N+2)+E(N)-2E(N+1)
\label{deltaN}
\end{equation}
where $E(N)$ is the $N$-body ground state energy,
\begin{equation}
     {\tilde \Delta}(N+2)<0
\end{equation}
signals a bound pair in the ground state with $N+2$ particles and
$|{\tilde \Delta}|$ is interpreted as the binding energy of the
pair. Pairing was found under various conditions by several
authors\cite{balseiro},\cite{gubernatis}, and among the geometries
considered the $4\times 4$ system\cite{fop} is one of the most
relevant for studying the pairing instability close to the
antiferromagnetic phase.

 Below, we derive an analytic  theory of pairing interactions,
 and report   several detailed case studies where the formulas
 are  validated by comparison with the numerical data. The results
 support general qualitative criteria for pairing induced by the on-site repulsion only.
  This  shows that our approach successfully predicts the formation of bound pairs, and
  explains why other ingredients like strong off-site interactions
are needed  in other geometries. The analytic approach is also
needed to predict what happens for large systems and in the
thermodynamic limit. Increasing the cluster size the computed pair
binding energies show a rapid decrease, and several authors on the
basis of the numerical data  consider pairing in the Hubbard model
as a size effect.  Unfortunately, the number of configurations
grows in a prohibitive  way with the cluster size\cite{size} and
numerical data currently available on 4$\times$4 or even
6$\times$6 clusters cannot  provide reliable extrapolations to the
bulk limit. In Section \ref{section3.4} we show that we understand
the trend analytically very well, that much larger cells (at least
30$\times$30)  are needed to estimate the asymptotic behaviour and
that we have reasons to believe that pairing with a reduced but
substantial binding energy persists in the full plane.

\subsection{$W=0$ Pairing in  Cu-O  Clusters}
\label{section2.1}

  In this Section, we illustrate the concept of $W=0$ pairs
and the way they become bound states, by using examples with a
geometry relevant for the cuprates.  Our starting point is the
three-band Hubbard Hamiltonian
\begin{equation}
H=K+W+W_{\mathrm{off-site}}
\label{tbhh}
\end{equation}
where
$$
K= t\sum_{\bra ij\ket,\s}(p^{\dag}_{j\s}d_{i\s}+{\mathrm
h.c.})+t_{pp}\sum_{\bra
jj'\ket,\s}p^{\dag}_{j\s}p_{j'\s}+\ve_{d}\sum_{i,\s}n_{i\s}+
\ve_{p}\sum_{j,\s} n_{j\s}
$$
and
$$
W=U_{d}\sum_{i}n_{i\ua}n_{i\da}+U_{p}
\sum_{j}n_{j\ua}n_{j\da},\;\;\;\;\;
W_{\mathrm{off-site}}=U_{pd}\sum_{\bra
ij\ket,\s\s'}n_{i\s}n_{j\s'}.
$$
Here, $p_{j}$ ($d_{i}$) are fermionic operators that destroy holes
at the Oxygen (Copper) ions labelled $j$ ($i$) and
$n=p^{\dag}p\; (\; =d^{\dag}d) $ is the number operator.
$\bra ij\ket$ refers to pairs of nearest neighbors $i$ (Copper)
and $j$ (Oxygen) sites. The hopping terms correspond to the
hybridization between nearest neighbors Cu and O atoms, and are
roughly proportional to the overlap between localized Wannier
orbitals.

The parameters $U_{d}$ and $U_{p}$ are positive constants that
represent the repulsion between holes when they are at the same
Copper ($d$) and Oxygen ($p$) orbitals, respectively. $U_{pd}$ has
a similar meaning, \textit{i.e.}, it corresponds to the Coulombic
repulsion when two holes occupy two adiacent Cu-O sites.  The
on-site energies $\ve_{p}$ and $\ve_{d}$ refer to the occupied
orbitals of Oxygen and Copper. In the strong coupling limit, and
with one particle per unit cell, this model reduces to the spin
Heisenberg Model with a superexchange antiferromagnetic
coupling\cite{emery2},\cite{fulde}.

From a band structure calculation and by best fitting the results of
\textit{ab initio} calculations\cite{hyber} one can
roughly estimate the actual values of the parameters in the
Hamiltonian of Eq. (\ref{tbhh}). Our preferred set is (in eV):
$\ve_{p}-\ve_{d}= 3.5$,  $t=1.3$, $t_{pp}=-0.65 $, $U_{d}=5.3$,
$U_{p}=6$ and, most probably,  $U_{pd}<1.2$.

As Cini and Balzarotti\cite{cb1} pointed out, highly symmetric
clusters  possess 2-holes singlet eigenstates of $H$  which do not
feel the on-site repulsion $W$; such eigenstates were called
\textit{$W=0$ pairs} and play a crucial role for pairing.  In
order to have $W=0$ solutions, the clusters must possess the full
$C_{4v}$ (square) symmetry, and must be centered around a Cu site.
Group arguments are central to our approach and we show the
Characters of the $C_{4v}$ Group in Table \ref{c4vtable}.
\begin{table}
\begin{center}
\begin{tabular}{|c|c|c|c|c|c|c|}
\hline $C_{4v}$ & $\mathbf{1}$ & $C_{2}$ &
$C^{(+)}_{4},\,C^{(-)}_{4}$ & $\s_{x},\,\s_{y}$ &
$\s_{+},\,\s_{-}$ & Symmetry \\
\hline
$A_{1}$ & 1 & 1 & 1 & 1 & 1 & $x^{2}+y^{2}$\\
\hline
$A_{2}$ & 1 & 1 & 1 & -1 & -1 & $(x/y)-(y/x)$\\
\hline
$B_{1}$ & 1 & 1 & -1 & 1 & -1 & $x^{2}-y^{2}$\\
\hline
$B_{2}$ & 1 & 1 & -1 & -1 & 1 & $xy$ \\
\hline
$E$ & 2 & -2 & 0 & 0 & 0 & $(x,y)$ \\
\hline
\end{tabular}
\caption{Character table of the $C_{4v}$ symmetry Group. Here
$\mathbf{1}$ denotes the identity,  $C_{2}$ the  180 degrees
rotation,  $C_{4}^{(+)},\;C_{4}^{(-)}$ the counterclockwise and
clockwise 90 degrees rotation, $\s_{x},\;\s_{y}$ the reflection
with respect to the $y=0$ and $x=0$ axis and $\s_{+},\;\s_{-}$ the
reflection with respect to the $x=y$ and $x=-y$ diagonals. The
$C_{4v}$ symmetry Group has 4 one-dimensional irreducible
representations (irreps) $A_{1}$, $A_{2}$, $B_{1}$, $B_{2}$ and 1
two-dimensional irrep $E$. In the last column it is shown how each
of them transforms under $C_{4v}$.} \label{c4vtable}
\end{center}
\end{table}
The symmetry requirements   are so stringent that
 clusters with such properties had not been studied
previously. In particular, the present discussion does not apply
to the  geometries, like those examined by Hirsch \textit{et
al.}\cite{scalah},\cite{scalah2} and Balseiro \textit{et
al.}\cite{balseiro}, which are {\em forbidden} from our viewpoint.
The Cu$_{4}$O$_{4}$ geometry considered by Ogata and
Shiba\cite{ogata} has the $C_{4v}$ symmetry, but lacks the central
Cu, and therefore it is {\em forbidden}. Those built from
degenerate one-body levels are not $W=0$ pairs; they feel the
on-site repulsion $U_{p}$ on the Oxygen sites, and yield
$\tilde{\D}<0$ only if $U_{p}$ is very
small\cite{cb1},\cite{martin}.

 Note that we are using the same $t$ for all
Cu-O  bonds, while some authors use other conventions. For
instance, F. C. Zhang and T. M. Rice\cite{zr} use an alternating
sign prescription for $t$, which may be obtained by changing the
sign of all the O orbitals in the horizontal lines containing Cu
ions. The two pictures are related by a gauge transformation,
under which the orbital symmetry labels $A_{1}$ and $B_{1}$ are
interchanged. Some of the symmetry related information is gauge
dependent and unobservable, while some is physical (e.g.
degeneracies are).

The reason why symmetry is so basic for our mechanism is that the
$W=0$ two-body eigenstates arise when two holes occupy degenerate
one-body levels.  When such degenerate states are partially filled
in the many-body ground state,   bound pairs can form. The {\em
interacting ground state} must be described in terms of a
many-body configuration mixing, but the presence of the $W=0$ pair
imparts to the many-body state a special character; it can be
described in terms of a bound pair moving on a closed-shell
background. This is the main point that we want to make, and it is
deeply related to the symmetry quantum numbers.

\begin{figure}[!ht]
\centering \resizebox{10cm}{!}{
\includegraphics{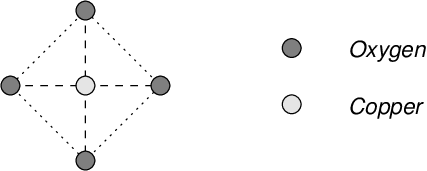}
} \caption{{\footnotesize The topology of the  $\cuoq$ cluster.}}
\label{cup:schemetto-cuo4}
\end{figure}

The smallest square Cu-O cluster, $\cu$,  see Fig.
\ref{cup:schemetto-cuo4}, is also the simplest where  $W=0$
pairing occurs. The Hamiltonian reads:
\begin{equation}
H_{\rm{CuO}_{4}}=  t \sum_{i\s}( d^{\dag}_{ \s}p_{ i\s}+
p_{  i\s}^{\dag}d_{ \s})+ t_{pp} \sum_{<ij>,\s} p^{\dag}_{
i,\s}p_{ j,\s}+U(n^{(d)}_{  \ua} n^{(d)}_{
\da}+\sum_{i}n^{ (p)}_{  i \ua}n^{(p)}_{  i\da} ) .
\label{sch}
\end{equation}
The one-body levels and their symmetry labels are reported in
Table \ref{o-bel}.
\begin{table}
\begin{center}
\begin{tabular}{|c|c|c|c|c|}
\hline
       $\varepsilon_{A_{1}}$ &
       $\varepsilon_{E_{x}}$ &
       $\varepsilon_{E_{y}}$ &
       $\varepsilon_{B_{1}}$ &
       $\varepsilon_{A_{1}'}$ \\
\hline
       $\t-\sqrt{4+\t^2}$ &
       $0$ &
       $0$ &
       $-2\t$ &
       $\t+\sqrt{4+\t^2}$ \\
\hline
\end{tabular}
\caption{One-body levels of the CuO$_{4}$ cluster in units of $t$ as a
function of the dimensionless parameter $\t\equiv t_{pp}/t$.
} \label{o-bel}
\end{center}
\end{table}
Labelling the Oxygen atomic sites by the numbers $1$, $2$, $3$,
$4$, the diagonalizing creation operator are given by
\begin{eqnarray}\label{onebody}
\Ccr{\ebid_{y} \sigma} & = & \frac{1}{\sqrt{2}} \left( \Pcr{2
\sigma} - \Pcr{4 \sigma}
\right) \nonumber \\
\Ccr{\ebid_{x} \sigma} & = & \frac{1}{\sqrt{2}} \left( \Pcr{1
\sigma} - \Pcr{3 \sigma}
\right)  \nonumber \\
\Ccr{\buno \sigma} & = & \frac{1}{2} \left( \Pcr{1 \sigma} -
\Pcr{2 \sigma} + \Pcr{3 \sigma} - \Pcr{4 \sigma}
\right)  \nonumber\\
\Ccr{\auno \sigma}(1) & = & \frac{1}{\alpha^{2}_{+}+4} \left( \alpha_{+}
\Dcr{\sigma} + \Pcr{1 \sigma} + \Pcr{2 \sigma} + \Pcr{3 \sigma} +
\Pcr{4 \sigma}
\right) \nonumber \\
\Ccr{\auno \sigma}(2) & = & \frac{1}{\a^{2}_{-}+4} \left( \a_{-}
\Dcr{\sigma} + \Pcr{1 \sigma} + \Pcr{2 \sigma} + \Pcr{3 \sigma} +
\Pcr{4 \sigma}
\right)
\end{eqnarray}
where $\alpha$ and $\beta$ depend on $\tau=t_{pp}/t$ as follows
$$
 \a_{\pm} = \frac{4\,\left( \pm 1 \pm  {\tau }^2 +
         \tau \,{\sqrt{4 + {\tau }^2}} \right) }{\pm 5\,
        \tau  \pm 2\,{\tau }^3 + {\sqrt{4 + {\tau }^2}} +
       2\,{\tau }^2\,{\sqrt{4 + {\tau }^2}}}.
$$

\begin{figure}[!ht]\label{many}
\centering \resizebox{10cm}{!}{
\includegraphics{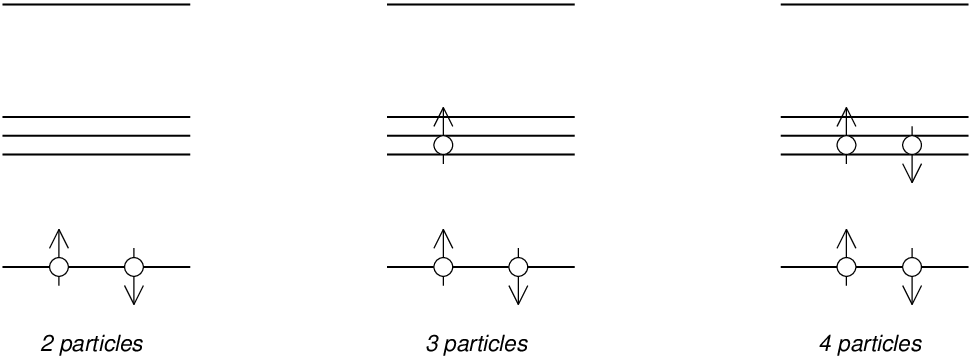}
} \caption{{\footnotesize Many body states for the $\cuoq$
cluster. For each number of particle, it has been reported the
component of
highest weigth of the ground state.}}
\end{figure}

Let us build a 4-hole state in the CuO$_{4}$  cluster in the
non-interacting limit, according to the \textit{aufbau }
principle, see Fig.2. The first two holes go into a bonding level
of $A_{1}$ symmetry; this is a totally symmetric ($^{1}A_{1}$)
pair. For negative $\t$, the other two holes go into a non-bonding
level of $E(x,y)$ symmetry, which contains 4 spin-orbital states.
The Pauli principle allows $\left( \begin{array}{c}4 \\
2 \\ \end{array} \right) =6$ different pair-states. The irrep multiplication
table allows for labelling them according with their space symmetry:
$E\otimes E = A_{1}\bigoplus A_{2}\bigoplus B_{1}\bigoplus B_{2}$. It
is also straightforward to verify that $A_{2}$ is a spin-triplet,
$^{3}A_{2}$, while the remaining irreps are spin-singlets,
$^{1}A_{1}$, $^{1}B_{1}$ and $^{1}B_{2}$. From Eq. (\ref{onebody}) one
readily realize that the $B_{2}$ singlet operator
\begin{equation}
\label{coppia}
 \Bcr{\bdue}  =  \displaystyle \frac{1}{\sqrt{2}} \left(
            \Ccr{\ebid_{x} \ua} \Ccr{\ebid_{y} \da} +
            \Ccr{\ebid_{y} \ua} \Ccr{\ebid_{x} \da}
            \right)
\end{equation}
is  a $W=0$ pair (no double occupation). Note that $^{1}B_{2}$ is the
symmetry label of the pair wave function in the gauge we are using,
and must not be confused with the symmetry of the order parameter.

\begin{figure}[!ht]
\centering
\includegraphics{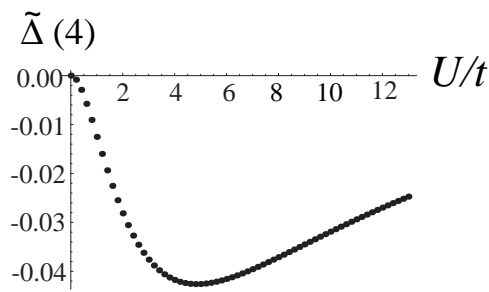}
\caption{$\tilde{\D}(4)$ (in $t$ units) as a function of $U/t$.
The maximum binding occurs at $U \sim 5 t$ where
$\tilde{\D}(4) \approx -0.042\;t$.  For $U> 34.77\; t$ (not shown),
$\tilde{\D}(4)$ becomes positive and pairing disappears.}
\label{fig01}
\end{figure}

To first order in perturbation theory, the 4-body
singlet state of $B_{2}$ symmetry is degenerate with the $A_{2}$
triplet; Hund's rule would have
predicted a $^{3}A_{2}$ ground state. However, the true ground
state turns out to be singlet, for reasons that we shall study below.
The numerical results on the $\cu$ cluster show that $\tilde{\D}(4)$
is negative for $0>\t > -0.5$ and that its minimum value
occurs at $\t=0$, when the non-bonding orbitals $B_{1}$ and
$E$ become degenerate: the paired interacting ground state is also
degenerate in this particular case. A symmetry analysis of this
accidental degeneracy is postponed to the next Section.  In Fig.
\ref{fig01} we plot $\tilde{\D}(4)$ for $\t=0$,
$\ve_{p}-\ve_{d}=0$, $U_{pd}=0$ and $U_{p}=U_{d}=U$.
$\tilde{\D}(4)$ has a minimum at $U\approx 5\; t$ and it is
negative  when $0< U < 34.77\;t$. We emphasize that
$\tilde{\D}(4)$ becomes positive for large values of $U/t$ and
hence pairing disappears in the strong coupling regime. In the
present problem $U$ must exceed several tens of times $t$ before
the asymptotic \textit{strong coupling  regime} sets in. A
perturbation theory will strictly apply at \textit{weak coupling}
where the second derivative of the curve is negative. However,
qualitatively a weak coupling approach is rewarding in all the
physically interesting range of parameters. The sign of
$\tilde{\D}$ depends on $U$ and $\t$ and its magnitude is unlike any of the
input parameters; below we show that this new energy scale comes out from an
interference between electron-hole exchanges of different symmetries.

On the other hand, at positive $\t$'s the $B_{1}$ non-bonding level is pushed
below the degenerate one and $\tilde{\D}(4)$ becomes large and
positive (at $\t=+0.65$, $\tilde{\D}(4)=0.53$ eV).

Next, we discuss the dependence of $\tilde{\D}(4)$ on the other
parameters\cite{cb1}. If we decrease $\ve_{p}$, $\tilde{\D}(4)$
decreases because this makes the system more polarizable. The $\ve_{p}$
dependence when all the other parameter are kept fixed and $U_{pd}=0$ is
almost linear down to $\ve_{p}=0$. According to
Ref.\cite{balseiro} positive $U_{pd}$ values do not spoil the mechanism,
and tend to be synergic with it. Indeed, values of $U_{pd}>0.6$ eV give
negative $\tilde{\D}(4)$ values even for $\t=-0.65$ eV (in the range
0.2-1.2 eV considered in Ref.\cite{cb1} $\tilde{\D}(4)$ is a monotonically
decreasing function of $U_{pd}$). Finally, we have numerically studied
how the distortions effect $\tilde{\D}(4)$. We have found that any lowering of
the symmetry is reflected by a corresponding increase of $\tilde{\D}(4)$\cite{cb1}.

In the $\cu$ cluster we need four holes to have a paired ground
state. Hence, the total hole concentration is $\r_{h}=0.8$. This
value is too large by a factor of 2 with respect to the
experimentally observed $\r_{h}\simeq 0.4$ of the optimally doped
systems. It is important to realize that these undesirable
features are peculiar of the  prototype CuO$_{4}$ cluster, and
already disappear in Cu$_{5}$O$_{4}$, the next larger cluster of
the same symmetry. In fact, four holes are still sufficient to
reach degenerate states, but $\r_{h}\sim 0.44$ is much closer to
the experimental value. We have performed numerical explorations
in other fully symmetric clusters like Cu$_{5}$O$_{4}$ and
Cu$_{5}$O$_{16}$ and we have found negative values of
$\tilde{\D}(4)$, of the order of few meV, using physical
parameters. In all the allowed clusters up to 21 atoms, the lowest
one-hole level belongs to $A_{1}$ symmetry, and the next $E$ level
yields the $W=0$ pair. The interactions produce a non-degenerate
$^{1}B_{2}$ 4-hole ground state \textit{having the same symmetry
as the $W=0$ pair}. The interested reader may see
Refs.\cite{cb34},\cite{EPJB2000} for the details.
 Below, in Section \ref{section3} we
show\cite{cb5} that the $\tilde{\D}(4)<0$ arises from an effective
attractive interaction between the holes of the $W=0$ pair; the
same interaction  is repulsive for triplet pairs.

\subsection{Symmetry of the $W=0$ pairs: a general theorem}
\label{section2.2}

Since the mechanism depends on symmetry in such a fundamental way,
we must refine the Group Theory analysis. We discovered a powerful and
elegant criterion\cite{ijmp00} to construct  all the $W=0$ pair
eigenstates on a given lattice $\L$ by using projection operators.

Let ${\cal G}_{0}$ be the the symmetry Group
of the non-interacting Hubbard Hamiltonian
$K=\sum_{\bra ij\ket,\s}t_{ij}c^{\dag}_{j\s}c_{i\s}$. We assume that no
degeneracy between one-body eigenstates is accidental, hence ${\cal
G}_{0}$ must contain enough operations to justify all such
degeneracies. Let us label  each  one-body eigenstate of
$H=K+ W$, $W=\sum_{i\in\L}U_{i}
n_{i\ua}n_{i\da}$, with an  irreducible representation
(irrep) of ${\cal G}_{0}$.

\textit{Definition}.  An irrep $\eta$ is
represented in the one-body spectrum of $H$ if at
least one of the one-body levels belongs to $\eta$.\\

Let ${\cal E}$ be the set of the irreps of ${\cal G}_{0}$ which are represented in
the one-body spectrum of $H$. Let $|\q\rangle$ be
a two-body eigenstate of the non-interacting Hamiltonian with spin
$S^{z}=0$ and $P^{(\eta)}$ the projection operator on the irrep
$\eta$. We wish to prove the\\

\begin{figure}[!ht]
\centering \resizebox{12cm}{!}{
\includegraphics{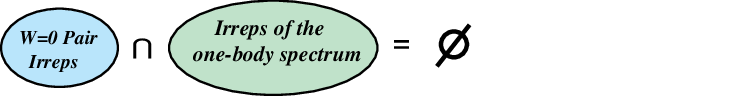}
} \caption{{\footnotesize Schematic of the $W=0$ theorem.}}
\label{inter}
\end{figure}

\underline{\textit{$W=0$ Theorem.}} -
Any nonvanishing projection of $|\q\rangle$ on an irrep
\underline{not} contained in ${\cal E}$, is an eigenstate of
$H$ with no double occupancy. The singlet component of this state is a
$W=0$ pair. Conversely, any pair belonging to an irrep represented in the
spectrum must have non-vanishing $W$ expectation
value, see Fig. (\ref{inter}):

\begin{equation}
\label{teoremino}
\eta \notin  {\cal E} \Leftrightarrow W P^{(\eta)}|\q\rangle=0.
\end{equation}

\textit{Remark}: Suppose we perform a gauge change in
$H$ such that ${\cal G}_{0}$ is preserved;
clearly, a $W=0$ pair goes to another  $W=0$ pair. Thus, the
theorem implies a distinction between symmetry types which is
gauge-independent.

\textit{Proof}: Let us consider a two-body state of
opposite spins belonging to the irrep $\eta$ of ${\cal G}_{0}$:
$$
|\q^{(\eta)}\rangle=\sum_{ij\in\L}\q^{(\eta)}(i,j)
c^{\dag}_{i\ua}c^{\dag}_{j\da}|0\rangle.
$$
Then we have
$$
n_{i\ua}n_{i\da}|\q^{(\eta)}\rangle=
\q^{(\eta)}(i,i)c^{\dag}_{i\ua}c^{\dag}_{i\da}|0\rangle\equiv
\q^{(\eta)}(i,i)|i\ua,i\da\rangle.
$$
We define $P^{(\eta)}$ as the projection operator on the irrep $\eta$.
Since
$$
P^{(\eta)}\sum_{i\in\L}\q^{(\eta)}(i,i)|i\ua,i\da\rangle=
\sum_{i\in\L}\q^{(\eta)}(i,i)|i\ua,i\da\rangle,
$$
if $P^{(\eta)}|i\ua,i\da\rangle=0\;\forall i\in\L$, then
$\q^{(\eta)}(i,i)=0\;\forall i\in\L$. It is worth to note that this
condition is satisfied if and only if
$P^{(\eta)}|i\s\rangle=0\;\forall i\in\L$, where
$|i\s\rangle=c^{\dag}_{i\s}|0\ket$. Now it is always possible to
write $ |i\s\rangle$ in the form  $|i\s\rangle=\sum_{\eta\in {\cal
E}}c^{(\eta)}(i) |\eta_{\s}\rangle$ where $|\eta_{\s}\rangle$ is the
one-body eigenstate of $H$ with spin $\s$
belonging to the irrep $\eta$.  Hence, if $\eta'\notin {\cal E}$,
$P^{(\eta')}|i\s\rangle=0$ and so $P^{(\eta')}|i\ua,i\da\rangle=0$.
Therefore, if $|\q^{(\eta)}\ket$ is a two-hole singlet eigenstate of
the kinetic term and $\eta\notin {\cal E}$, then it is also an
eigenstate of $W$ with vanishing eigenvalue, that means a $W=0$
pair.

Q.E.D.\\

\bigskip
The content of the theorem is schematically represented by Fig. (\ref{inter}).

We already know that CuO$_{4}$  is a good example of this theorem.
Indeed, the irrep $B_{2}$ of the $W=0$ pair is not represented in
the spectrum  (see Table \ref{o-bel}); $A_{2}$ is also not
represented, but it yields no two-body states. This possibility is
admitted by Eq. (\ref{teoremino}).

The  particular case of CuO$_{4}$  with  $\t=0$ is of special
interest as an illustration. There is an \textit{accidental}
degeneracy between $E(x,y)$ and $B_{1}$ orbitals and therefore  a
three-times degenerate one-body level exists. With 4 interacting fermions,
pairing occurs in two ways, as  $A_{1}$ and $B_{2}$
are both degenerate ground states. This agrees with the theorem;
in fact, the extra degeneracy cannot be explained in terms of
$C_{4v}$, whose irreps have dimension 2 at most. For $\t=0$
any permutation of the four Oxygen sites is actually a symmetry
and therefore  ${\cal G}_{0}$ is enlarged to $S_{4}$ (the group of
permutations of four objects). $S_{4}$ has the irreducible
representations ${\cal A}_{1}$ (total-symmetric), ${\cal B}_{2}$
(total-antisymmetric), ${\cal E}$ (self-dual), ${\cal T}_{1}$ and
its dual ${\cal T}_{2}$, of dimensions 1, 1, 2, 3 and 3,
respectively. These irreps break in $C_{4v}$ as follows
$$
{\cal A}_{1}=A_{1},\;\;\;{\cal T}_{1}=B_{1}\oplus E,\;\;\; {\cal
T}_{2}=A_{2}\oplus E,\;\;\; {\cal B}_{2}=B_{2},\;\;\; {\cal
E}=A_{1}\oplus B_{2}\;.
$$
Labelling the one-body levels with the irreps of $S_{4}$, one
finds that ${\cal E}$ is not contained in the spectrum and thus
yields $W=0$ pairs:
\begin{equation}
^{1}\cebid\ =\ ^{1}A_{1}\ \oplus\ ^{1}B_{2} \;.
\end{equation}
The explicit expression of the corresponding pair-creation operators
is
$$
\Bcr{{\auno}}  =  \displaystyle \frac{2}{\sqrt{3}}
            \Ccr{\buno \ua} \Ccr{\buno \da} +
            \frac{1}{\sqrt{3}} \left(
            \Ccr{\ebid_{x} \ua} \Ccr{\ebid_{x} \da} +
            \Ccr{\ebid_{y} \ua} \Ccr{\ebid_{y} \da}
            \right)
$$
for the total-symmetric pair and Eq. (\ref{coppia}) for the $B_{2}$
component.

In Section \ref{section4.3} we shall show use the theorem with a
${\cal G}_{0}$ that includes symmetries which are not
isometries. We shall see that this theorem puts useful
restrictions on the many-body ground state symmetry.
The complete characterization of the symmetry of $W=0$ pairs requires
the knowledge of ${\cal G}_{0}$. A partial use of the
theorem is possible if one  does not know
${\cal G}_{0}$ but knows a Subgroup ${\cal G}^{<}_{0}$.
It is then still granted that any pair belonging to an irrep of
${\cal G}^{<}_{0}$ not
represented in the spectrum has the $W=0$ property.  On the other
hand, accidental degeneracies occur with a Subgroup of ${\cal
G}_{0}$, and by mixing degenerate pairs belonging to irreps represented
in the spectrum one can find $W=0$ pairs also there.

The theorem tells us that the \textit{bigger} is ${\cal G}_{0}$, the
\textit{larger} is the
number of $W=0$ pairs. Indeed, for a given system, the number
of one-body eigenvectors is fixed, while the number and the dimension
of the irreps grow with the order of the group.
Therefore, also the number of irreps not represented in ${\cal E}$
grows, meaning more $W=0$ pairs. At the same time, a big
``non-interacting'' symmetry Group ${\cal G}_{0}$ grants large level
degeneracies. As we have seen, this fact allows the
existence of $W=0$ pairs formed by degenerate orbitals. In the next
Section we show that an anomalously low effective repulsion takes place
in the interacting system when such levels are on the Fermi surface
and that in a certain region of the parameters space
this leads to pairing.

\section{Theory of Pairing in repulsive  Hubbard Models}
\label{section3}

In Section \ref{section2} we have shown that $W=0$ pairs may
lead to a paired ground state in fully symmetric $C_{4v}$ clusters
centered around a Cu ion. In this Section we extend the theory to the
full plane, and show that such pairs provide the natural explanation
of the pairing instability of the Hubbard Model. By a novel canonical
tranformation approach, we shall calculate the effective interaction
$W_{\mathrm{eff}}$ between two holes added to the ground state of the
repulsive Hubbard Model. Furthermore, we shall show that $W_{\mathrm{eff}}$
is attractive in the $W=0$ pair-symmetry channels. The method is a particularly
efficient way to perform the configuration interaction calculation. It is
based on the symmetry and sheds light on the origin of pairing in the
CuO$_{2}$ plane. It is worth noticing that $W=0$ pairs emerge from
symmetry alone and hence remain $W=0$ for any coupling strength
$U$. This is the reason why weak coupling expansions provide often
good approximations at intermediate coupling as well, as observed by
several authors\cite{metzner2},\cite{fridman},\cite{galan}.

 \subsection{$W=0$ Pairs in the Planar Hubbard Model}
\label{section3.1}

The CuO$_{2}$ planes of cuprates have a large symmetry group which
contains the space group ${\cal G}=T\otimes C_{4v}$, where $T$ is the
group of translations and $\otimes$ stands for the semidirect product.
However, the Optimal Group ${\cal G}_{0}$ is much larger than
${\cal G}$\cite{optimal}. For this reason, instead of using the $W=0$
theorem, it is easier to follow the simple route of projecting a single
determinantal state on the irreps of  $C_{4v}$\cite{EPJB1999},\cite{SSC1999}.
This partial use of the symmetry still gives enough solutions to
demonstrate pairing.

Let us focus on $W=0$ pairs with vanishing quasimomentum. Omitting band
indices, we denote by
\begin{equation}
|d[{\mathbf k}]\ket=c^{\dagger}_{{\mathbf k},\ua}
c^{\dagger}_{-{\mathbf k},\da}|0\ket
\label{detercoop}
\end{equation}
a two-hole determinantal state, where ${\mathbf k}$
and $-{\mathbf k}$ label degenerate one-body ei\-gen\-fun\-ctions of $K$.
The combination $|d[{\mathbf k}]\ket+|d[-{\mathbf k}]\ket$
is singlet and $|d[{\mathbf k}]\ket-|d[-{\mathbf k}]\ket$ is triplet.
We introduce the determinants
$|d[R{\mathbf k}]\ket\equiv|d[{\mathbf k}_{R}]\ket,\;R\in C_{4v}$, and
the projected states
\begin{equation}
|\Phi_{\eta}\left[ {\mathbf k}\right] \ket=\frac{1}{\sqrt{8}}{\sum_{R\in
C_{4v}}}\chi
^{\left( \eta \right) }\left( R\right)|d[{\mathbf k}_{R}]\ket
\label{detrproj}
\end{equation}
where $\chi^{\left( \eta \right) }(R)$ is the character of the
operation $R$ in the irrep $\eta$.

In the Three Band Model we showed that $W=0$ pairs are obtained by
projecting onto the irreps $A_{2}$ and $B_{2}$\cite{obhm} (which are not
represented in the one-body spectrum, as it should). These irreps
differ by those obtained in the fully symmetric clusters, where
$W=0$ pairs had $A_{1}$ and $B_{2}$ symmetry\cite{cbs1}. The reason
for this change is a twofold size effect. On one hand, $A_{1}$ pairs
have the $W=0$ property only in clusters having the topology of a cross
(whose symmetry group is $S_{4}$ rather than $C_{4v}$), but do not
generalize as such to the full plane (when the symmetry is lowered to
$C_{4v}$). On the other hand, the small clusters admit no $W=0$ solutions of
$A_{2}$ symmetry if only degenerate states are used.

We recall that a necessary condition for pairing in clusters is that
the least bound holes form a $W=0$ pair, and this dictates conditions
on the occupation number. In the full plane, however, $W=0$ pairs exist
at the Fermi level for any filling. We also observe that the  $W=0$ pairs
obtained with the above procedure are not \textit{all} the
possible $W=0$ pairs in the One and Three Bands Hubbard models, since
the $W=0$ theorem has not been fully exploited. However, it can be
shown that they are the only $W=0$ pairs with holes in degenerate
one-body levels and with vanishing quasimomentum.

\subsection{Canonical Transformation}
\label{section3.2}

In this Section we intend to study the effective interaction
$W_{\rm eff}$ between the particles forming a $W=0$ pair added to the
$N$-body interacting ground state $|\F_{U}(N) \rangle$. Since the two
extra particles cannot interact directly (by definition of $W=0$ pair),
their effective interaction comes out from virtual electron-hole exchanges,
and in principle can be attractive.

Many configurations contribute to the interacting $(N+2)$-body ground
state $|\F_{U}(N+2) \rangle$ and we need a complete set $\cal{S}$ to
expand it exactly; as long as it is complete, however, we can design
$\cal{S}$ as we please. We can take the non-interacting  $N$-body
Fermi \textit{sphere} $|\F_{0}(N)\rangle$ as our vacuum and build the
complete set in terms of excitations over it.
In the subspace with vanishing spin $z$ component, the simplest states
that enter the configuration mixing are those obtained from
$|\F_{0}(N)\rangle$ by creating two extra holes\cite{e-hsymmetry} over
it; we denote with $|m\ket$ these states. Similarly, along with the pair
$m$ states, we introduce the  4-body $\alpha $ states, obtained from
$|\F_{0}(N)\ket$ by creating $2$ holes and 1 electron-hole (e-h) pair.
Then, $\mathcal{S}$ includes the 6-body $\beta$ states  having $2$
holes and 2 e-h pairs, and so on. We are using Greek indices for the
configurations containing the electron-hole pairs, which here are playing
largely the same role as phonons in the Cooper theory. By means of
the complet set $\mathcal{S}$ we now expand the interacting ground state
\begin{equation}
|\F_{U}(N+2)\ket={\sum_{m}}a_{m}|m\ket+{\sum_{\alpha }}
a_{\alpha }|\alpha \ket+{\sum_{\beta }}a_{\beta }
|\beta \ket+....
\label{lungo}
\end{equation}
and set up the Schr\"{o}dinger equation
$$
H|\F_{U}(N+2)\ket=E|\F_{U}(N+2)\ket.
\label{seq}
$$
We emphasize that Eq. (\ref{lungo}) is a configuration interaction, \textit{not a
perturbative expansion}. When  the number $N$  of holes in the system is
such that $|\F_{0}(N)\ket$ is a single non-degenerate determinant
(the Fermi surface is totally filled), the expansion (\ref{lungo}) for the
interacting ground state is unique and we can unambiguously define and
calculate the effective interaction $W_{\rm eff}$. This is done  by a
canonical transformation\cite{EPJB2000},\cite{EPJB1999},\cite{SSC1999}
of the many-body Hubbard Hamiltonian. We consider the effects of the
operators $K$ and $W$ on the terms of $|\F_{U}(N+2)\ket$. Choosing
the $m$, $\a$, $\b$, $\ldots$ states to be eigenstates of the kinetic
operator $K$, we have
$$
K|m\ket=E_{m}|m\ket,  \;\;\;\;
K|\a\ket=E_{\a}|\a\ket,  \;\;\;\;
K|\b\ket=E_{\b}|\b\ket,  \;\;\;\;\ldots.
$$
The Hubbard interaction $W$ can create or destroy up to 2 e-h pairs.
Therefore, its action on an $m$ state yields
$$
W|m\ket={\sum_{m^{\prime }}}
W_{m^{\prime },m}|m^{\prime }\ket+
\sum_{\alpha}W_{\alpha,m}|\alpha\ket +
\sum_{\beta}W_{\beta ,m}|\beta\ket ,
$$
on an  $\a$ states yields
$$
W|\alpha\ket={\sum_{m}}W_{m,\alpha }|m\ket +
\sum_{\alpha^{\prime }}W_{\alpha ^{\prime },\alpha }|\alpha^{\prime
}\ket+
\sum_{\beta} W_{\beta ,\alpha}|\beta\ket+
\sum_{\g}W_{\g\a}|\g\ket ,
$$
and so on. In this way we obtain an algebraic system for the coefficients
of the configuration interaction of Eq. (\ref{lungo}). In order to test
the instability of the Fermi liquid towards pairing, it is sufficient to
study the amplitudes $a_{m}$ of the $m$ states. In the weak coupling limit
this can be done by truncating the expansion (\ref{lungo}) to the $\a$
states. As we have shown\cite{SSC1999}, the inclusion of the $\b,\;\g,\ldots$
states produces a $E$-dependent renormalization of the matrix elements
of higher order in $W$, leaving the structure of the equations unaltered.

By taking a linear combination of the $\a$ states in such a way that
$$
(K+W)_{\a,\a'}=\d_{\a\a'}E'_{\a}
$$
the algebraic system reduces to
\begin{equation}
\left[ E_{m}-E\right]
a_{m}+{\sum_{m^{\prime }}}
W_{m,m^{\prime }}a_{m^{\prime}}
+{\sum_{\alpha }}W_{m,\alpha }a_{\alpha }=0
\label{reson1}
\end{equation}
$$
\left[ E'_{\alpha }-E\right] a_{\alpha }+
{\sum_{m^{\prime }}}W_{\alpha ,m^{\prime }}a_{m^{\prime}}=0.
$$
Solving for $a_{\alpha }$ and substituting back in Eq. (\ref{reson1}),
we end up with an equation for the dressed pair
$|\q\rangle\equiv \sum_{m}a_{m}|m\rangle$.
The effective Schr\"{o}dinger equation for the pair reads
\begin{equation}
\left(K+W+S[E]\right) |\q\ket \equiv H_{\rm pair} |\q\ket =E|\q\ket
\label{can1}
\end{equation}
where
\begin{equation}
(S[E])_{m,m'}=-\sum_{\a}\frac{W_{m,\a}W_{\a,m'}}{E'_{\a}-E}.
\label{fscat}
\end{equation}
is the scattering operator. The matrix elements $W_{m,m'}$ in Eq. (\ref{can1})
may be written as the sum of two terms representing the direct interaction
$W^{(d)}_{m,m'}$ between the particles forming the pair and the first-order
self-energy $W_{m}$:
$$
W_{m,m'}=W^{(d)}_{m,m'}+\d_{m,m'}W_{m}.
$$
Similarly, in $S[E]$ we may recognize two different contributions; one
is the true effective interaction $W_{\rm eff}$, while the other is the
forward scattering term $F$
$$
S_{m,m'}=(W_{\rm eff})_{m,m'}+F_{m}\d_{m,m'} \,.
$$
The first-order self-energy and the forward scattering term are diagonal
in the indices $m$ and $m'$.  $W_{m}$ and $F_{m}$ renormalize the
non-interacting energy $E_{m}$ of the $m$ states:
$$
E_{m}\ra E^{(R)}_{m}=E_{m}+W_{m}+F_{m}.
$$
Eq. (\ref{can1}) has the form of a Schr\"odinger equation with
eigenvalue $E$ for the added pair with the interaction $W^{(d)}+W_{\rm eff}$.
Here, the $W=0$ pairs are special because $W^{(d)}$ vanishes. We interpret
$a_{m}$ as the wave function of the dressed pair, which is acted upon by an
effective Hamiltonian $H_{\rm pair}$. This way of looking at Eq. (\ref{can1}) is
perfectly consistent, despite the presence of  the many-body eigenvalue $E$.
Indeed, if the interaction is attractive and produces bound states, the
spectrum contains discrete states below the threshold of the continuum
(two-electron Fermi energy). This is a clear-cut criterion for pairing, which
is exact in principle. The threshold is given by
$$
E^{(R)}_{F}\equiv \min_{\{m\}}[ E^{(R)}_{m}(E)],
$$
and contains all the pairwise interactions except those between the particles
in the pair. $E^{(R)}_{F}$ must be determined once Eq. (\ref{can1}) has been solved
(since $F$ depends on the solution). The ground state energy $E$ may be conveniently
written as $E^{(R)}_{F}+\D$. $\D<0$ indicates a Cooper-like instability of the
normal Fermi liquid and its magnitude represents the binding energy of the pair.

It is worth noticing that in  principle the canonical transformation
is exact and it is not limited to the weak coupling regime.
In the numerical calculations, however, some approximation is needed. In
practice, we shall compute the bare quantities, that is, we shall neglect
the 6-body and higher excitations in the calculation of $W_{\rm eff}$ and $F$.
This is a resonable approximation if we compute small corrections to a
Fermi liquid background, and the exact numerical results obtained in
small clusters suggest that this is the case, see next Section.

\subsection{Pairing in the CuO$_{4}$ Cluster}
\label{section3.3}

We defined the pairing energy for a cluster with $N+2$ particles by
introducing the quantity ${\tilde \Delta}(N+2)=E(N+2)+E(N)-2E(N+1)$.
At weak coupling we have shown in Ref.\cite{cb5}
that $\tilde{\D}$ agrees with the binding energy $\D$ obtained from
many-body perturbation theory. Here, we consider again the
CuO$_{4}$ cluster; we intend
to test our canonical transformation by comparing the analytic
results with the exact diagonalization data obtained in Section
\ref{section2}.

The CuO$_{4}$ cluster can host a $W=0$ pair of $B_{2}$ symmetry.
At weak coupling, we can ignore the effect of the renormalizations
in Eq. (\ref{fscat}). Furthermore, we can discard the inter-shell
interactions. Indeed, the one-body levels of finite clusters are
widely separated and the dominant contribution to the effective
interation comes from the intra-shell interaction. Accordingly, in
Eq. (\ref{fscat}) we set the $m=m'=B_{2}$. The one-body levels and
their symmetry labels are reported in  Table \ref{o-bel}.

The effective interaction in the $B_{2}$ symmetry channel is
$$
W_{\rm eff}^{(B_{2})}\equiv \sum_{\a}
\frac{W_{B_{2},\a}W_{\a,B_{2}}}{E_{\a}-E}=-\frac{U^{2}}{32}
\left[\frac{2}{\varepsilon_{B_{1}}+\varepsilon_{A_{1}}-E} -
\frac{1}{\varepsilon_{A'_{1}}+\varepsilon_{A_{1}}-E}\right].
$$

\begin{figure}[!ht]
\centering
\includegraphics{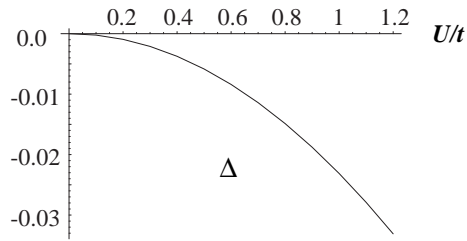}
\caption{Trend of $\D$ versus $U/t$ in units of $t$. }
\label{fig03}
\end{figure}

The binding energy is obtained by expressing the lowest eigenvalue $E$
of $H_{\rm pair}$ as $E=E_{F}^{(R)}+\D$. The renormalized two-hole
Fermi energy $E_{F}^{(R)}$ coincides with the bare one at
this level of accuracy, \textit{i.e.}, $E_{F}^{(R)}=2\varepsilon_{A_{1}}$.
The results are shown in Fig. \ref{fig03} for $\t=0$. We can see that
$\D$ is negative and hence the renormalization approach predicts pairing,
at least at weak coupling. We also observe that the order of
magnitude of $|\Delta|$ is $10^{-2}t$, which is not comparable to any
of the $U$ and $t$ input parameters. The reason
is that the interaction, which vanishes identically for the {\em
bare} $W=0$ pairs, remains \textit{dynamically} small for the dressed
quasiparticles. This suggests that a \textit{weak coupling} theory may be
useful to study the pairing force, despite the fact that $U$ is not
small compared to $t$.

Next, we intend to compare $\D$ with the quantity ${\tilde \Delta}(4)$
obtained from exact diagonalizations, see Fig. \ref{fig01}.
At weak coupling the agreement is excellent. However, $|\Delta|$ is
$\sim  2$ times greater than $|\tilde{\D}(4)|$ for $U/t\simeq 1$. This means
that the inter-shell interactions and the renormalizations of the $\a$-state
energies have an important weight in determining the right value of
$\Delta$. However, what is comfortable is that the analytic approach
predicts the right trend of the binding energy. In the next Section
we shall use the canonical transformation to study larger and more
physical systems.

\subsection{Pairing in the CuO$_{2}$ Plane: Numerical Results}
\label{section3.4}

Using the analytic expression for the effective interaction in the
full plane, we have performed exploratory numerical estimates of $\Delta$
by working on supercells of $N_{\L}\times N_{\L}$ cells, with periodic
boundary conditions. For the sake of simplicity, we have neglected the minor
contributions from the higher bands and considered the dominant intra-band
processes, in which empty states belong to the bonding band. We have
solved the problem in a virtually exact way for $N_{\L}$ up to 40.
Several supercell calculations had been reported to
date\cite{dagrev}, but no conclusive evidence of a pairing instability
was reached due to the difficulty of dealing with size effects.

First, we have investigated triplet pairing but, as in the clusters, we have
found a repulsive effective interaction. On the contrary, $W=0$
singlets show pairing, in line with our previous findings in
small clusters\cite{cb1},\cite{cb34},\cite{cb5}.
Since screening excitations are explicitly accounted for in the Hamiltonian,
it is likely that $U$ is a bare (unscreened) quantity, which justify large
values. A stronger interaction causes smaller pairs and speeds up
convergence within attainable supercell sizes. In Table \ref{epj99II},
we report the results for $^{1}B_{2}$ pairs  with $\ve_{F}=-1.3$ eV
[half filling corresponds to $\ve_{F}=-1.384$ eV and we have used as input
data the set of current parameters already used for clusters, that is
(in eV) $t=1.3$, $\ve_{p}=3.5$, $\ve_{d}=0$, $U_{p}=12.5$, $U_{d}=11.2$].

\begin{table}
\begin{center}
\begin{tabular}{|c|c|c|c|c|}
  \hline
$N_{\L}$ & $ \r=N/N^{2}_{\L}$ & $ -\Delta$ (meV) &
$V_{\mathrm{eff}}$ (eV)  & $-\Delta _{\mathrm{asympt}}$ (meV)   \\
  \hline
18 & 1.13 & 121.9 & 7.8 & 41.6   \\
  \hline
20 & 1.16 & 42.2 & 5.0 & 9.0   \\
  \hline
24 & 1.14 & 59.7 & 7.0 & 28.9   \\
  \hline
30 & 1.14 & 56.0 & 5.7 & 13.2   \\
  \hline
40 & 1.16 & 30.5 & 6.6 & 23.4 \\
  \hline
\end{tabular}
\caption{Binding energy of $^{1}B_{2}$ pairs in supercells.}
\label{epj99II}
\end{center}
\end{table}

With supercell sizes $N_{\L}>40$ calculations become hard. Thus, we
need a simple solvable model in supercells and in the infinite plane
to make reliable extrapolations of numerical results. To this end,
we introduce an Average Effective Interaction (AEI)
$-V_{\mathrm{eff}},\;V_{\mathrm{eff}}>0$, which is constant
for all the empty states $\mathbf{k}$ and $\mathbf{k}^{\prime }$.
For any $N_{\L}$, $V_{\mathrm{eff}}$ is implicitly defined from
$$
\frac{8}{V_{\mathrm{eff}}}=\frac{1}{N_{\L}^{2}}\sum_{\bf k}
\frac{\th(\ve_{\bf k}-\ve_{F})}
{2(\ve_{\bf k}-\ve_{F})+|\D|},
$$
(the factor 8 comes from the projection onto the irrep $B_{2}$).
Although $|\D|$ decreases monotonically by increasing $N_{\L}$,
$V_{\mathrm{eff}}$ remains fairly stable around $6\div 7$ eV, see
Table \ref{epj99II}. The relatively mild $N_{\L}$ dependence of
$V_{\mathrm{eff}}$ supports the use of the AEI to extrapolate the
results to the thermodynamic limit. The asymptotic value $\lim_{N_{\L}\ra\inf}|\D|
\equiv \Delta_{\mathrm{asympt}}$ can then be obtained from
$$
\frac{8}{V_{\mathrm{eff}}}=\int_{\ve_{F}}^{0}\frac{d\ve \rho \left
(\ve\right)
}{2\left( \ve -\ve_{F}\right) +|\Delta_{\mathrm{asympt}}|}  ,
$$
where $\rho$ is the density of states.

The results for the $^{1}A_{2}$ pairs are seen to lead to bound states
as well, with comparable $\Delta$ values\cite{EPJB1999}; the trend with
doping is opposite, however, and the binding energy is nearly closing at
$\ve_{F}=-1.1$ eV.

Although the three-band Hubbard Model is an idealization of the
strongly correlated CuO$_{2}$ planes, it is interesting to observe
that evidence of mixed $(s+id)$ symmetry for the pairing state has
been amply reported in angle-resolved photoemission
studies\cite{shen1},\cite{lynch}.

\section{The Doped Hubbard Antiferromagnet}
\label{section4}

The canonical transformation described in Section \ref{section3.2}
relies on the uniqueness of the non-interacting ground state
$|\Phi_{0}(N)\ket$. $|\Phi_{0}(N)\ket$ is certainly unique if the
Fermi surface is totally filled and it can be written as a Slater
determinant. However, we are also particularly
interested in the doped Hubbard antiferromagnet, and the antiferromagnetic
ground state occurs at half filling, \textit{not} in a filled-shell situation.

We want to study the doped antiferromagnet since there are strong
indications that the Fermi liquid is unstable towards pairing  near
half filling; they come from diagrammatic approaches\cite{bickers},
renormalization group techniques\cite{zanchi},\cite{metzner} and also
cluster diagonalizations\cite{fop}. Therefore
exact results on the half filled Hubbard Model may be relevant to
antiferromagnetism and to the mechanism of the superconducting instability
as well. The Lieb theorem\cite{lieb} on the ground state
spin-degeneracy of the half-filled Hubbard model ensures that for
any bipartite lattice $\L={\cal A}\cup{\cal B}$, with $|{\cal A}|=|{\cal B}|$,
the ground state is unique for any interaction strenght $U$. Thus, we
can use the canonical transformation with $|\Phi_{0}(N)\ket=\lim_{U\ra 0^{+}}
|\F_{U}(N)\ket$ provided the number of holes $N$ equals the number of sites
$|\L|$. Remarkably, the $W=0$ pairs are also essential tools to deal with
the antiferromagnetic ground state solution. Below, we present two exact results.
First, we obtain the ground state $|\Phi_{U}(|\L|)\ket$ on the so called
\textit{complete bipartite graph} (CBG) for arbitrary but finite $U$.
Then, we consider the square lattice and we
resolve the degeneracy of the non-interacting ground-state multiplet
at half filling. This latter result will then be used to study the
pairing mechanism in the $4\times 4$ Hubbard model.

\subsection{Half-Filled Hubbard Model on the Complete Bipartite Graph}
\label{section4.1}

\begin{figure}[!ht]
\centering \resizebox{10cm}{!}{
\includegraphics{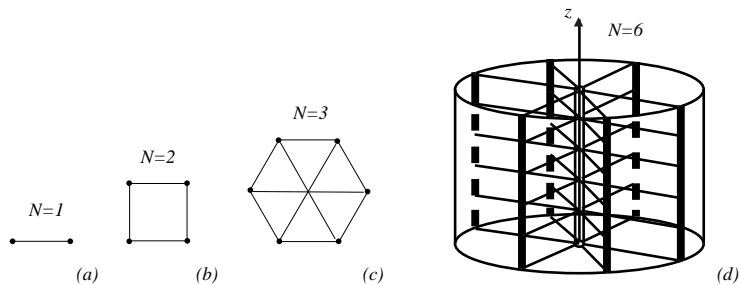}
} \caption{{\footnotesize Physical representation of the CBG for
    $N=1$ ($a$) and $N=2$ ($b$) which is equivalent to a one dimensional ring
        of length $L=2,\;4$ respectively. For $N=3$ ($c$) we have the (1,1)
        nanotube of smallest length and periodic boundary conditions. For
        $N=6$ ($d$) we present  the {\em gedanken} device described in the
        text.}}
\label{clust}
\end{figure}

The Complete Bipartite Graph (CBG) $\L={\cal A}\cup{\cal B}$ has
bonds connecting any element of ${\cal A}$ with all the elements
of ${\cal B}$.
In Fig. \ref{clust} we have drawn a few examples of finite-size systems.
For $N=1$ and $N=2$ the model is equivalent to a one dimensional ring
of length $L=2,\;4$ respectively. For $N=3$ we have what can be
understood as a prototype, (1,1)
{\em nanotube} model, the one of smallest length $L=1$, with periodic
boundary conditions. For
general $N$, one can conceive a {\em gedanken} device, like the one
illustrated  pictorially for $N=6$ in
Fig. \ref{clust}.$d$. The $N$ vertical lines represent a
realization of the  ${\cal A}$ sublattice while the  ${\cal B}$
sublattice is mimicked by the central object. The radial tracks in the
Figure represent conducting paths linking
each ${\cal A}$ site to each ${\cal B}$ site according to the topology
of our model.

Our solution is an example of \textit{antiferromagnetic} ground state
in a model of itinerant electrons; it may provide useful hints
about  antiferromagnetism  outside the strong coupling regime
(where the Hubbard model can be mapped onto the Heisenberg model).

The model is invariant under an arbitrary
permutation of the ${\cal A}$-sites and/or of the ${\cal B}$-sites.
In the symmetric case $|{\cal A}|=|{\cal B}|$
there is an additional $Z_{2}$
symmetry because of the ${\cal A} \leftrightarrow {\cal B}$ exchange.
In what follows we shall focus on the symmetric case and we call $\cN$
the number of sites in each sublattice, $|\L|=2\cN$.

The one-body spectrum has three different eigenvalues
$\ve_{g}=-t$, $\ve_{0}=0$ and $\ve_{\bar{g}}=t$ with degeneracy
$d_{g}=1$, $d_{0}=2\cN-2$ and $d_{\bar{g}}=1$
respectively. We use the convention $t>0$ so that $\ve_{g}$ is the
lowest level and we shall call ${\cal S}_{hf}$ the set of zero-energy
one-body eigenstates, $|{\cal S}_{hf}|=2\cN-2$. The zero-energy one-body
orbitals can be visualized by a simple argument. Consider
any pair $i,\;j$, with $i \neq j$, of sites belonging to the same
sublattice, say ${\cal A}$, and a wavefunction $\psi_{i,j}$  taking the
values 1 and -1 on the
pair and 0 elsewhere in ${\cal A}$ and in  ${\cal B}$. It is evident
that $\psi_{i,j}$ belongs to ${\cal S}_{hf}$. Operating on  $\psi_{i,j}$
by the permutations of $S_{\cN}$ we can generate a (non-orthogonal)
basis of $\cN-1$ eigenfunctions vanishing in ${\cal B}$; further,
by means of the $Z_{2}$ symmetry,  we obtain the remaining orbitals
of ${\cal S}_{hf}$, which vanish on ${\cal A}$. This exercise shows
that the group considered above justifies the ($2\cN-2$)-fold degeneracy
of the one-body spectrum.

We denote by $g$ ($\bar{g}$) the operator which annihilates a
particle in the eigenstate with energy $\ve_{g}$ ($\ve_{\bar{g}}$).
Then, the kinetic term $K$ can be written as
$$
K=-t\sum_{\s}(g^{\dag}_{\s}g_{\s}-\bar{g}^{\dag}_{\s}\bar{g}_{\s})\;.
$$
Next, we introduce the one-body eigenstate $a^{\dag}_{i}|0\ket$ of
${\cal S}_{hf}$ with vanishing amplitudes on the ${\cal B}$ sublattice.
Similarly, $b^{\dag}_{i}|0\ket$ has vanishing amplitude on the ${\cal A}$
sublattice and
therefore the $(2\cN-2)$-body state
\begin{equation}
|\F_{AF}^{(\s)}\ket=a^{\dag}_{1\s}\ldots a^{\dag}_{\cN-1\s}
b^{\dag}_{1\bar{\s}}\ldots b^{\dag}_{\cN-1\bar{\s}}|0\ket,\;\;\;\;
\bar{\s}=-\s
\label{detaf}
\end{equation}
is an eigenstate of $K$ and of $W$ with vanishing
eigenvalue. In
the following we shall use the wording \textit{$W=0$ state} to denote any
eigenstate of $H$ in the kernel of $W$. It
is worth noticing that by mapping the ${\cal A}$-sites
onto the ${\cal B}$-sites and \textit{viceversa}, $|\F_{AF}\ket$ retains
its form except for a spin-flip; we  call this property the {\it
antiferromagnetic property} for obvious reasons.

The state $|\F_{AF}\ket$ plays a crucial role in building the exact
ground state at half filling. We first observe that $|\F_{AF}\ket$
has non-vanishing projection onto all the total spin subspaces
$S=0,\ldots,\cN -1$. Let us denote with
$|\F_{AF}^{0,0}\ket$ the singlet component and with $|\F_{AF}^{1,m}\ket$,
$m=0,\pm 1$, the triplet component. We further introduce the
short-hand notation
$|g^{0}\ket\equiv g^{\dag}_{\ua}g^{\dag}_{\da}|0\ket$ and
$|\bar{g}^{0}\ket\equiv
\bar{g}^{\dag}_{\ua}\bar{g}^{\dag}_{\da}|0\ket$
for the two-body singlets that one obtains from the lowest and
the highest energy orbitals $g$ and $\bar{g}$, and
$|[g\bar{g}]^{1,m}\ket$, $m=0,\pm 1$,
for the corresponding triplet. Then, one can prove\cite{sc} that
the interacting ground state $|\F_{U}(|\L|)\ket$ can be written as
$$
|\F_{U}(|\L|)\ket=\left[\g_{g}|g^{0}\ket+\g_{\bar{g}}|\bar{g}^{0}\ket\right]
\otimes|\F_{AF}^{0,0}\ket+
\g_{0}\sum_{m=-1}^{1}
(-)^{m}|[g\bar{g}]^{1,m}\ket\otimes|\F_{AF}^{1,-m}\ket\;,
$$
where $(\g_{g},\g_{\bar{g}},\g_{0})$ is the lowest energy eigenvector
of a $3\times 3$ Hermitean matrix\cite{sc}.

We observe that $|\F_{U}(|\L|)\ket$ is an eigenstate of
the total number operator $\hat{n}^{a}+\hat{n}^{b}$ of particles in
the shell ${\cal S}_{hf}$, despite the fact that
$[\hat{n}^{a}+\hat{n}^{b},H]\neq 0$. Such a remarkable
property (\textit{shell population rigidity}) relies on the
cancellation of those scattering amplitudes that do not preserve the number of
particles in ${\cal S}_{hf}$. We have shown\cite{sc} that this
cancellation
takes place provided  in ${\cal S}_{hf}$ the $(2\cN-2)$-body state is a
$W=0$ state.
We have also found that the ground state energy $E(|\L|)$ is negative for
any value of the repulsion $U$; qualitatively, we may say that the
particles
manage to  avoid the double occupation very effectively. Furthermore,
$E(|\L|)$ is a monotonically increasing function of $U$ and $\cN$ due to
the existence of non-trivial correlations even for large $\cN$. The
nature
of these correlations has been investigated by computing the expectation
value of the  repulsion. We have shown that for any finite $\cN$ there
is a
critical value of $U$ yielding maximum repulsion. Even more
interesting is the magnetic order of the ground state. Due to the
$S_{\cN}\otimes S_{\cN}\otimes Z_{2}$ symmetry, the spin-spin
correlation function $G_{\rm spin}(i,j)\equiv
\bra\F_{U}(|\L|)|S^{z}_{i}S^{z}_{j}|\F_{U}(|\L|)\ket$ can be written as
$$
G_{\rm spin}(i,j)=\left\{\begin{array}{ll}
G_{0} & i=j \\
G_{\rm on} & i\in{\cal A}\;(i\in{\cal B})\;{\rm and}\;
j\in{\cal A}\;(j\in{\cal B})\\
G_{\rm off} & i\in{\cal A}\;(i\in{\cal B})\;{\rm and}\;
j\in{\cal B}\;(j\in{\cal A})
\end{array}\right.
$$

\begin{figure}[!ht]
\centering
\includegraphics{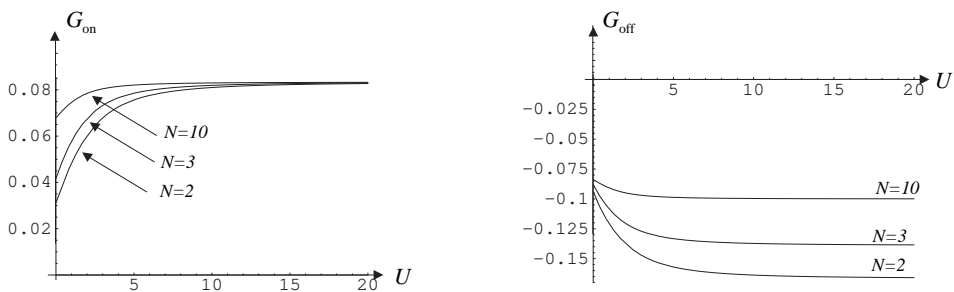}
\caption{$a$) $G_{\rm on}$ versus $U$ in the range $0\leq U\leq
         20$ for three different values of the number of sites
         $\cN=2,\;3,\;10$. $b$) $G_{\rm off}$ versus $U$ in the range
     $0\leq U\leq 20$ for three different values of the number of sites
         $\cN=2,\;3,\;10$.
    The hopping parameter has been chosen to be $t=1$ in both cases.}
\label{fig04}
\end{figure}

In Fig. \ref{fig04} we report the trend of $G_{\rm on}$ and
$G_{\rm off}$ versus $U$ for three different values of
$\cN=2,\;3,\;10$. According to the Shen-Qiu-Tian theorem\cite{sqt},
$G_{\rm on}$ is always larger than zero while $G_{\rm
off}$ is always negative. Next, we consider the ground state average
of the square of the staggered magnetization operator
\begin{equation}
m^{2}_{AF}\equiv
\frac{1}{|\L|}\bra\F_{U}|[\sum_{i\in\L}\e(i)S^{z}_{i}]^{2}
|\F_{U}\ket,\;\;\;\;
\e(i)=1,\;-1\;{\rm for}\;\; i\in{\cal A},\;{\cal B}\;.
\label{sqtfor}
\end{equation}
The Shen-Qiu-Tian theorem implies that each term in the expansion
of Eq. (\ref{sqtfor}) is non-negative. We emphasize, however, that
for $|{\cal A}|=|{\cal B}|$ this does not imply that $m^{2}_{AF}$
is an extensive quantity! Remarkably,
in the CBG $m^{2}_{AF}=G_{0}+(\cN-1)G_{\rm on}-
\cN G_{\rm off}$ is extensive for any value of the on-site
repulsion $U$ and provide the first example
of antiferromagnetic ground state in a model of itinerant electrons.

\subsection{Half-Filled Hubbard Model on the Square Lattice for $U=0^{+}$}
\label{section4.2}

We consider the Hubbard Hamiltonian on the square lattice
\begin{equation}
H=K+W=
\frac{t}{2}\sum_{\bra{\mathbf r},{\mathbf r}'\ket\s}
(c^{\dag}_{{\mathbf r}\s}c_{{\mathbf r}'\s}+{\rm h.c.})+U\sum_{{\mathbf
r}}
n_{{\mathbf r}\ua}n_{{\mathbf r}\da}
\label{hhafgs}
\end{equation}
with ${\mathbf r}=(i_{x},i_{y})$, $i_{x},i_{y}=1,\ldots,N_{\L}$ and
the sum on $\bra{\mathbf r},{\mathbf r}'\ket$ is over
the pairs of nearest neighbors sites. The point symmetry is
$C_{4v}$; besides, $H$ is
invariant under the  commutative Group of translations ${\mathbf
T}$ and hence the space Group   ${\mathbf G}={\mathbf
T} \otimes C_{4v} $; $\otimes$ means the semidirect product.
In terms of the Fourier expanded fermion operators (periodic boundary
conditions) $c_{{\mathbf k}}=\frac{1}{N_{\L}}\sum_{{\mathbf r}}
e^{i {\mathbf k}\cdot {\mathbf r}}c_{{\mathbf r}}$, we have
$K=\sum_{{\mathbf k}}\ve_{{\mathbf k}}c^{\dag}_{{\mathbf
k}\s}c_{{\mathbf k}\s}$
with $\ve_{{\mathbf k}}=2t[\cos k_{x}+\cos k_{y}]$.
Then, the one-body plane wave state $c^{\dag}_{{\mathbf k}\s}|0\ket
\equiv|{\mathbf k}\s\ket$ is an  eigenstate of $K$.

In this Section we build the exact ground state of the Hubbard
Hamiltonian (\ref{hhafgs}) at half filling  and weak coupling for a
general \textit{even} $N_{\L}$.
Once a unique non-interacting ground state is determined, one can use
the non-perturbative canonical transformation to test the instability
of the system towards pairing; this will be done in the next Section
for $N_{\L}=4$.

The starting point is the following property of the number operator
$n_{{\mathbf r}}=c^{\dag}_{{\mathbf r}}c_{{\mathbf r}}$
(for the moment we omit the spin index).

\underline{\textit{Theorem}}:   Let ${\cal S}$ be an arbitrary set
of plane-wave eigenstates $\{|{\mathbf k}_{i}\ket\}$ of $K$ and
$(n_{{\mathbf r}})_{ij}=\bra {\mathbf k}_{i}|n_{{\mathbf r}}|
{\mathbf k}_{j}\ket=\frac{1}{N_{\L}^{2}}
e^{i({\mathbf k}_{i}-{\mathbf k}_{j})\cdot{\mathbf r}}$  the matrix of
$n_{{\mathbf r}}$ in ${\cal S}$.
This  matrix  has eigenvalues $\l_{1}=\frac{|{\cal S}|}{N_{\L}^{2}}$
and $\l_{2}= \ldots =\l_{|{\cal S}|}=0$.

Note that  $|{\cal S}|\leq N_{\L}^{2}$; if $|{\cal S}|= N_{\L}^{2}$
the set is complete, like the  set of all orbitals, and the theorem is
trivial (a particle on site ${\mathbf r}$ is the
$n_{{\mathbf r}}$ eigenstate with eigenvalue
1). Otherwise, if $|{\cal S}|< N_{\L}^{2}$, the theorem is an immediate
consequence of the fact\cite{jop2001} that
\begin{equation}
{\rm det}|(n_{{\mathbf r}})_{ij}-\l\d_{ij}|=(-\l)^{|{\cal S}|-1}
(\frac{|{\cal S}|}{N_{\L}^{2}}-\l),\;\;\;\forall {\mathbf r}.
\label{det}
\end{equation}

Let ${\cal S}_{hf}$ denote the set (or shell) of the ${\mathbf k}$ wave
vectors
such that $\ve_{{\mathbf k}}=0$.
At half filling ($N^{2}_{\L}$ particles) for $U=0$ the ${\cal S}_{hf}$
shell
is half occupied, while all ${\mathbf k}$ orbitals such that
$\ve_{{\mathbf k}}<0$ are filled. The ${\mathbf k}$ vectors of
${\cal S}_{hf}$ lie on the square having
vertices $(\pm\pi,0)$ and $(0,\pm\pi)$;
one  readily realizes that the dimension of the
set ${\cal S}_{hf}$ is $|{\cal S}_{hf}|=2N_{\L}-2$. Since $N_{\L}$ is
even
and $H$ commutes with the total spin operators,
at half filling every ground state of $K$ is represented in
the $S^{z}=0$ subspace. Thus, $K$ has
$\left(\begin{array}{c} 2N_{\L}-2 \\ N_{\L}-1
\end{array}\right)^{2}$   degenerate unperturbed ground state
configurations with $S^{z}=0$. Most of the degeneracy is removed in
first-order by $W$. We shall be
able to single out the unique  ground state of $H$
by exploiting the Lieb theorem.

The first-order splitting of the degeneracy is obtained by diagonalizing
the $W$  matrix over the unperturbed basis; like in elementary
atomic physics, the filled shells just produce a constant shift of all
the eigenvalues and for the rest may be ignored in first-order
calculations. In other terms we consider the \textit{truncated  Hilbert
space ${\cal H}$} spanned by the \textit{states of $N_{\L}-1$ holes of each
spin in ${\cal S}_{hf}$}, and  we want the \textit{exact} ground state(s) of
$W$ in  ${\cal H}$; by construction ${\cal H}$ is in the kernel
of $K$, so the ground state of $W$ is the ground state of $H$ as
well. Since the lowest eigenvalue of $W$ is zero, it is evident that
any $W=0$ state in ${\cal H}$ is a ground state of $H$.

To diagonalize the \textit{local} operator $W$ in closed form
we need to set up a \textit{local}  basis set of one-body states. If
${\cal S}_{hf}$ were the complete set of plane-wave states
${\mathbf k}$, the new basis would be trivially obtained by a Fourier
transformation, but this is not the case. We introduce a set
$\{|\varphi_{\a}^{({\mathbf r})}\ket\}$ of orbitals such that
$n_{{\mathbf r}}$ is diagonal in this basis. The eigenvectors
$|\varphi_{\a}^{(0)}\ket$ of $n_{{\mathbf r}=0}$ and those
$|\varphi^{({\mathbf r})}_{\a}\ket$ of other sites ${\mathbf r}$ are
connected by translation and also by a unitary transformation,
or change of basis set. Picking ${\mathbf r}=\hat{e}_{l}$,  $l=x$ means
$\hat{e}_{l}=(1,0)$ or transfer by one step towards the right and $l=y$
means
$\hat{e}_{l}=(0,1)$ or transfer by one step upwards. The unitary
transformation
reads:
\begin{equation}
|\varphi^{(\hat{e}_{l})}_{\a}\rangle=
\sum_{\b=1}^{2N_{\L}-2}|\varphi_{\b}^{(0)}\rangle
\bra\varphi_{\b}^{(0)}|\varphi^{(\hat{e}_{l})}_{\a}\rangle
\equiv \sum_{\b=1}^{2N_{\L}-2}|\varphi_{\b}^{(0)}\rangle T_{l_{\b\a}}.
\label{transferT}
\end{equation}
The transfer matrix $T_{l}$  \textit{knows} all the translational and point
symmetry
of the system, and will turn out to be very special.

For large $N_{\L}$, to find $\{|\varphi_{\a}^{({\mathbf r})}\ket\}$
it is convenient to separate the ${\mathbf k}$'s of ${\cal S}_{hf}$ in
irreducible representations of the space Group $\mathbf{G}$$=
C_{4v}\otimes {\mathbf T}$. Choosing an arbitrary ${\mathbf k}\in {\cal
S}_{hf}$ with $k_{x}\geq k_{y}\geq 0$, the set of vectors $R_{i}{\mathbf
k}$,
where $R_{i}\in C_{4v}$, is a (translationally invariant) basis
for an irrep of $\mathbf{G}$. The high symmetry vectors $(0,\pi)$ and
$(\pi,0)$ always trasform among themselves and are the basis of the only
two-dimensional irrep of $\mathbf{G}$, which exists for any $N_{\L}$.
If $N_{\L}/2$ is even,  one also finds the high symmetry  wavevectors
${\mathbf k}=(\pm\pi/2,\pm\pi/2)$ which mix among themselves and yield
a four-dimensional irrep.  In general, the vectors $R_{i}{\mathbf k}$
are all
different, so all the other irreps of $\mathbf{G}$ have dimension 8,
the number of operations of the point Group $C_{4v}$.

Next, we show how to build our \textit{local} basis set and derive $W=0$
states  for each kind of irreps of $\mathbf{G}$. For illustration, we
shall first consider the case $N_{\L}=4$; then
${\cal S}_{hf}$ contains the bases of two irreps of $\mathbf{G}$,
of dimensions 2 and 4. The one with basis
${\mathbf k}_{A}=(\p,0),\;{\mathbf k}_{B}=(0,\p)$
breaks into $A_{1}\oplus B_{1}$ in $C_{4v }$.

The eigenstates of $(n_{{\mathbf r}=0})_{ij}=\bra
{\mathbf k}_{i}|n_{{\mathbf r}=0}|{\mathbf k}_{j}\ket$,
with $i,j=A,B$,
are $|\q''_{A_{1}}\ket=\frac{1}{\sqrt{2}}(|{\mathbf k}_{A}\ket+|{\mathbf
k}_{B}\ket)$
with $\lambda_{1}=1/8$
and $|\q''_{B_{1}}\ket=\frac{1}{\sqrt{2}}(|{\mathbf k}_{A}\ket-|{\mathbf
k}_{B}\ket)$
with $\lambda_{2}=0$.
Since under translation by a lattice step $T_{l}$ along the $l=x,y$
direction
$|{\mathbf k}\ket\ra e^{ik_{l}} |{\mathbf k}\ket$, using
Eq. (\ref{transferT}) one finds that
$|\q''_{A_{1}}\ket\leftrightarrow (-1)^{\th''_{l}}|\q''_{B_{1}}\ket$,
with $\th''_{x}=1,\;\th''_{y}=0$;  so $|\q''_{A_{1}}\ket$ has
vanishing amplitude on
a sublattice and $|\q''_{B_{1}}\ket$ on the other.
The two-body state $|\q''_{A_{1}}\ket_{\s}|\q''_{B_{1}}\ket_{-\s}$
has occupation for spin $\s$ but not for spin $-\s$ on the site
${\mathbf r}=0$;
under a lattice  step
translation it flips the spin and picks up a (-1) phase factor:
$|\q''_{A_{1}}\ket_{\s}|\q''_{B_{1}}\ket_{-\s}
\leftrightarrow
|\q''_{B_{1}}\ket_{\s}|\q''_{A_{1}}\ket_{-\s}$; therefore it has
double occupation nowhere and is a
$W=0$ state ($W=0$ pair\cite{SSC1999},\cite{EPJB1999}).

The 4-dimensional irrep with basis
${\mathbf k}_{1}=(\p/2,\p/2),\;{\mathbf k}_{2}=(-\p/2,\p/2),
\;{\mathbf k}_{3}=(\p/2,-\p/2)\;
{\mathbf k}_{4}=(-\p/2,-\p/2)$ breaks into $A_{1}\oplus B_{2}\oplus
E$ in $C_{4v}$; letting $I=1,2,3,4$ for the
irreps $A_{1},\;B_{2},\;E_{x},\;E_{y}$
respectively, we can  write down all the eigenvectors
of $(n_{{\mathbf r}=0})_{ij}=\bra {\mathbf k}_{i}|
n_{{\mathbf r}=0}|{\mathbf k}_{j}\ket$,
with $i,j=1,\ldots,4$, as
$|\q'_{I}\ket=\sum_{i=1}^{4}O'_{Ii}|{\mathbf k}_{i}\ket$, where $O'$ is
a 4$\times$4 orthogonal matrix.
The state with non-vanishing eigenvalue is again the $A_{1}$
eigenstate. After a little bit of algebra we have shown\cite{jop2001}
that
under
$T_{l}$ the subspace of  $A_{1}$ and $B_{2}$ symmetry is
exchanged with the one  of $E_{x}$ and $E_{y}$ symmetry. Thus
we can build a 4-body eigenstate of $W$ with vanishing
eigenvalue:
$|\q'_{A_{1}}\q'_{B_{2}}\ket_{\s}|\q'_{E_{x}}\q'_{E_{y}}\ket_{-\s}$.
As before under a lattice step translation
this state does not change its spatial
distribution but $\s\ra -\s$ without any phase factor:
$|\q'_{A_{1}}\q'_{B_{2}}\ket_{\s}|\q'_{E_{x}}\q'_{E_{y}}\ket_{-\s}
\leftrightarrow
|\q'_{E_{x}}\q'_{E_{y}}\ket_{\s}|\q'_{A_{1}}\q'_{B_{2}}\ket_{-\s}$.

Now we use these results to diagonalize $n_{{\mathbf r}=0}$
on the whole set ${\cal S}_{hf}$ (we could have done that directly by
diagonalizing $6 \times 6$ matrices but we wanted to show the general
method).
The eigenstate  of $n_{{\mathbf r}=0}$ with nonvanishing eigenvalue
always
belongs to  $A_{1}$.
The matrix $n_{{\mathbf r}}$
has  eigenvalues $3/8$ and (5 times) $0$, as predicted by Eq. (\ref{det}).
For ${\mathbf r}=0$ the eigenvector of occupation $3/8$ is
$|\varphi^{(0)}_{1}\rangle=
\frac{1}{\sqrt{3}}|\q''_{A_{1}}\ket+\sqrt{\frac{2}{3}}|\q'_{A_{1}}\ket$.
The other $A_{1}$
eigenstate of $n_{{\mathbf r}=0}$ has 0 eigenvalue and reads:
$|\varphi^{(0)}_{2}\rangle=
\sqrt{\frac{2}{3}}|\q''_{A_{1}}\ket-
\frac{1}{\sqrt{3}}|\q'_{A_{1}}\ket$.
The other eigenvectors, whose symmetry differs from $A_{1}$, are
$|\vf^{(0)}_{3}\ket=|\q'_{B_{2}}\ket$,
$|\vf^{(0)}_{4}\ket =|\q''_{B_{1}}\ket$,
$|\vf^{(0)}_{5}\ket =|\q'_{E_{x}}\ket$ and
$|\vf^{(0)}_{6}\ket=|\q'_{E_{y}}\ket$.
One finds\cite{jop2001} that the transfer matrices $T_{l}$ of
Eq. (\ref{transferT}) such that
$|\varphi^{(\hat{e}_{l})}_{I}\rangle
\equiv \sum_{J}|\varphi^{(0)}_{J}\rangle T_{l_{JI}}$,
are \textit{antiblock diagonal}.
Thus, the local basis at any site ${\bf r}$ splits into the subsets
${\cal S}_{a}=\{|\vf^{(0)}_{1}\rangle,|\vf^{(0)}_{2}\rangle,
|\vf^{(0)}_{3}\rangle\}$, and
${\cal S}_{b}=\{|\vf^{(0)}_{4}\rangle,|\vf^{(0)}_{5}\rangle,
|\vf^{(0)}_{6}\rangle\}$; a shift by a
lattice step sends members of ${\cal S}_{a}$ into linear combinations
of the members of  ${\cal S}_{b}$, and conversely.

Consider the 6-body eigenstate of $K$
$$
|\F_{AF}\ket_{\s}=
|\vf^{(0)}_{1}\vf^{(0)}_{2}\vf^{(0)}_{3}\rangle_{\s}
|\vf^{(0)}_{4}\vf^{(0)}_{5}\vf^{(0)}_{6}\rangle_{-\s} .
$$
In this state there is partial occupation of
site ${\mathbf r}=0$ with spin $\s$, but no double occupation. It turns
out that a
shift by a lattice step produces the transformation
$$
|\F_{AF}\ket_{\s}   \longleftrightarrow - |\F_{AF}\ket_{-\s}
$$
that is, a lattice step is equivalent to a spin flip, a feature that
we have already met in Section \ref{section4.1} (\textit{antiferromagnetic
property}). Since the spin-flipped state is
also free of double occupation, $|\F_{AF}\ket_{\s}$ is a $W=0$
eigenstate. A ground state which is a single determinant is a quite
unusual
property for an interacting model like this.

Note that $|\vf^{(0)}_{1}\vf^{(0)}_{2}\ket$ is
equivalent to $|\q''_{A_{1}}\q'_{A_{1}}\ket$, because this is just a
unitary transformation  of the $A_{1}$ wave functions; so
  $|\F_{AF}\ket_{\s}$ can also be
written in terms of the old  local orbitals (without any
mix of the  local states of different irreps of ${\mathbf G}$):
\begin{equation}
|\F_{AF}\ket_{\s}=|\q''_{A_{1}}\q'_{A_{1}}\q'_{B_{2}}\rangle_{\s}
|\q''_{B_{1}}\q'_{E_{x}}\q'_{E_{y}}\rangle_{-\s}.
\label{perirrep}
\end{equation}
This form of the ground state lends itself to be generalised to
arbitrary even $N_{\L}$, see Refs.\cite{SSC2001},\cite{jop2001}.

A few further remarks about $|\F_{AF}\ket_{\s}$ are in order.
(1) Introducing the projection operator
$P_{S}$  on the spin $S$ subspace, one finds that
$P_{S}|\F_{AF}\ket_{\s}\equiv|\F^{S}_{AF}\ket_{\s}\neq 0\; , \forall
S=0,\ldots,N_{\L}-1$.
Then, $_{\s}\bra \F_{AF}|W|\F_{AF}\ket_{\s}=\sum_{S=1}^{N_{\L}-1}\,
_{\s}\bra\F^{S}_{AF}|W|\F^{S}_{AF}
\ket_{\s}=0$, and this implies that there is at least one ground state
of $W$ in  ${\cal H}$ for each $S$.
The actual ground state of $H$ at weak coupling
is the singlet $|\F^{0}_{AF}\ket_{\s}$.  (2)
The \textit{existence} of this singlet $W=0$ ground state
is a direct consequence of the Lieb theorem\cite{lieb}. Indeed
the maximum spin state $|\F^{N_{\L}-1}_{AF}\ket_{\s}$ is trivially in
the kernel
of $W$; since the
ground state must be a singlet it should be an eigenvector of
$W$
with vanishing eigenvalue.  (3) The above results and
Lieb's theorem imply that
higher order effects split the ground state multiplet of $H$
and the singlet is lowest.  (4) The
Lieb  theorem makes no assumptions concerning the lattice
structure; adding the ingredient of the $\mathbf{G}$ symmetry we are able
to explicitly display the wave function at weak coupling.

Using the explicit form of $P_{S=0}$ one finds that
$P_{S=0}|\F_{AF}\ket_{\s}=-P_{S=0}|\F_{AF}\ket_{-\s}$. This
identity allows us to study how the singlet component transforms
under translations, reflections and rotations. In particular the {\em
antiferromagnetic property} tells us that the total momentum is
$K_{tot}=(0,0)$. To make contact with Ref.\cite{moreodag} we have
also determined how $|\F^{0}_{AF}\ket$ transforms under the $C_{4v}$
operations with respect to the center of an arbitrary plaquette.
It turns out\cite{jop2001} that it is even under reflections and
transforms as an
$s$ wave if $N_{\L}/2$ is even and as a $d$ wave if $N_{\L}/2$ is odd.

In the next Section we use these results, togheter with the
non-perturbative canonical transformation, to study the doped $4\times 4$
lattice at half filling. Since the non-interacting ground state at
half filling is now well known and unambiguously defined, the
expansion (\ref{lungo}) can be performed in a unique way.

\subsection{Pairing in the Doped Hubbard Antiferromagnet}
\label{section4.3}

The one-body spectrum of the 4$\times$4 Hubbard model
has 5 equally spaced levels, see Table \ref{4x4spectrum}.
The space Group $\mathbf{G}$ [containing the translations and the 8
$C_{4v}$ operations] can not explain the degeneracy 6, since
in the 4$\times$4 lattice the largest dimension of the irreps is 4.
As observed by previous authors\cite{bonca},\cite{poilb}, the
4$\times$4 lattice can be mapped into the $2\times 2\times 2\times 2$
hypercube since each pair of next to nearest neighbor sites
has two nearest neighbor sites in common. This implies
that $H$ is invariant under
a new and largest symmetry Group; let us call it ${\cal G}$.

Due to the importance of the symmetry in our configuration interaction
mechanism, we have explicitly calculated the character table of ${\cal G}$
taking into account an extra non-isometric symmetry operation\cite{EPJB2001}.
${\cal G}$ has 384 elements,  20 classes and hence 20 irreps whose
dimensionality fully justifies the degeneracies of Table \ref{4x4spectrum}.
\begin{table}
\begin{center}
\begin{tabular}{|c|c|c|}
\hline
    Energy&Irrep of ${\cal G}$ &Degeneracy \\
    \hline
    4 & $\tilde{B}_{2}$ & 1 \\
    \hline
    2 & $\L_{4}$ & 4  \\
    \hline
    0 & $\W_{4}$ & 6 \\
    \hline
    -2 & $\L_{1}$ & 4 \\
    \hline
    -4 & $A_{1}$ & 1 \\
    \hline
\end{tabular}
\caption{ One-body spectrum of the 4$\times$4 Hubbard model
for $t=-1$. We have used the notation introduced in
Ref.\cite{EPJB2001} in labelling the irreducible representations.}
\label{4x4spectrum}
\end{center}
\end{table}

As observed in Section \ref{section4.1} the  canonical
transformation applies when two holes are added to a non-degenerate
vacuum. To study the system at and close to half filling,
we have to use the results of Section \ref{section4.2}. In the following we
will solve the problem of two electrons added to the half filled system.

\subsubsection{$W=0$ Pairs}
\label{section4.3.1}

In order to study the $W=0$ pairing in the doped 4$\times$4
antiferromagnet, we consider $W=0$ pairs in the six-fold
degenerate one-body level [belonging to $\Omega_{4}$]. Exploiting
the $W=0$ theorem, we have found\cite{EPJB2001} $W=0$ pairs with
symmetry $\G_{1},\;\G_{2},\;\S_{2},\;\W_{1}$ and triplet pairs with
symmetry $\S_{3},\;\W_{2},\;\W_{3}$. Here, we are using the notation
of Ref.\cite{EPJB2001}; the irreps $\Omega$ have dimension 6, the
$\Sigma$'s  have dimension 3 and $\Gamma$'s have dimension 2.

Exact diagonalization results\cite{fop} show that for $U/t<3$ and
$16-2=14$ holes the ground state is sixfold degenerate. Below, we
use the canonical transformation and prove that the ground state
corresponds to an $\Omega_{1}$ $W=0$ \textit{electron} pair over the
half-filled system. For $U/t>3$ and the same number of holes a level
crossing takes place: the ground state is threefold degenerate and
contains a state with momentum $(0,0)$ and a doublet with momentum
$(\p,0)$ and $(0,\p)$. Again, the computed ground state can be assigned
to a $\S_{2}$ \textit{electron} pair over the half filled system\cite{EPJB2001}.

\subsubsection{Pairing mechanism}
\label{section4.3.2}

We consider the ground state of the $4\times 4$ Hubbard model with 14 holes;
aside from the 10 holes in the inner $A_{1}$ and $\Lambda_{1}$ shells
(see Table \ref{4x4spectrum}), the  outer $\W_{4}$ shell contains 4 holes.
We intend to show how pairing between two \textit{electrons} added to the
antiferromagnetic 16-holes ground state (half filling) comes out.
We use the antiferromagnetic ground state $|\F_{AF}^{S=0}\ket$ as the
non-interacting ground state of the configuration interaction
expansion (\ref{lungo}). With respect to this \textit{electron vacuum},
the $m$ states are now $W=0$ pairs of \textit{electrons} added to
$|\F_{AF}^{S=0}\ket$. In the $4\times 4$ lattice the one-body energy
levels are widely separated and the dominant interaction is between
electrons in the same shell. Therefore, we consider as $m$ states
only the $W=0$ electron pairs in the shell ${\cal S}_{hf}$,
\textit{i.e.}, those belonging to the irreps $\W_{1}$ and $\S_{2}$,
and we neglect the high-lying unoccupied orbitals.
Explicit calculations\cite{EPJB2001} show that the effective
interaction is attractive for both $\W_{1}$ and $\S_{2}$ $W=0$
electron pairs and that the corresponding binding energy is
$\D_{\W_{1}}=-61.9$ meV and $\D_{\S_{2}}=-60.7$ meV for $U=-t=1$ eV.
Therefore, the weak coupling ground state can be interpreted as an
$\W_{1}$ $W=0$ electron pair over the antiferromagnetic ground state.
This result agrees with exact diagonalization data\cite{fop}.

\section{Carbon Nanotubes and Triangular Cobalt Oxides}
\label{section5}

\subsection{Nanotubes}
\label{section5.1}

There is experimental evidence that the critical temperature $T_{c}$
in alkali-graphite intercalation compounds C$_{x}$M (where M is
a given alkali metal) grows as $x$ decreases\cite{belash}.
Under high-pressure, high metal concentration samples such as
C$_{6}$K, C$_{3}$K, C$_{4}$Na, C$_{3}$Na, C$_{2}$Na, C$_{2}$Li have
been synthesized; for C$_{2}$Na the value of $T_{c}$ is 5 K while for
C$_{2}$Li, $T_{c}$=1.9 K. Recently, Potassium\cite{bockrath2} and
Lithium\cite{gao} have been intercalated also in ropes of single-
and multi-wall carbon nanotubes (the highest metal concentration was
obtained with Lithium in C$_{2}$Li), and a net charge transfer
between the alkali-metals and the carbon atoms has been
predicted\cite{zhao}. The alkali-metals  cause little structural
deformation, but increase the filling of the original bands. Nanotubes
close to half filling  are deemed to be Luttinger liquids down to
milli-Kelvin temperatures\cite{balents},\cite{sol}.
Here\cite{PRB2002}, we use the Hubbard Hamiltonian $H$ on the wrapped
honeycomb lattice to represent the valence bands of single-wall carbon
nanotubes (SWNT) and we apply our symmetry-driven configuration
interaction pairing mechanism based on the existence of $W=0$ pairs.
We shall focus on armchair $(N,N)$ SWNT and show that the pair
binding energy grows as the number of electrons per C atom increases.
Furthermore, a stronger binding in nanotubes than in graphite sheets
is predicted which suggests a higher critical temperature in the former.
This is also supported by the measurements of a $T_{c}\approx 15$ K in
the 4 Angstrom SWNT by Tang \textit{et al.}\cite{tang}.

Any armchair SWNT has a bonding (-) and an antibonding (+) band, and
the Fermi line has $C_{2v}$ symmetry ($C_{6v}$ is the symmetry group
of the graphite sheet). Since these systems are usually doped with
\textit{electrons}, we take the Fermi level $\varepsilon_{F}$
to lie in the antibonding band. We use the $W=0$ theorem for
electrons with opposite quasimomentum and we find $W=0$ pairs belonging
to the pseudoscalar irrep $A_{2}$ of $C_{2v}$.
Let $(a,b)$ denote the basis of the Bravais lattice and
$u\left({\bf k},\z\right)$ the periodic part of
the Bloch function of quasi-momentum ${\bf k}$, with $\z =a,b$.
The singlet pair wavefunction reads\cite{PRB2002}
\begin{eqnarray}
\q^{[A_{2}]}_{\z_{1},\z_{2}}\left({\bf k}, {\bf R}_{1} , {\bf R}_{2}
\right) = \frac{1}{\sqrt{2}}
\sin \left( k_{x}(X_{1}-X_{2})\right) \times
\nonumber \\ \times
\left[u^{\ast}\left({\bf k},\z_{1}\right)
u^{\ast}\left(-{\bf k},\z_{2}\right)
e^{ik_{y}(Y_{1}-Y_{2})}-
u^{\ast}\left({\bf k},\z_{2}\right)
u^{\ast}\left(-{\bf k},\z_{1}\right)
e^{-ik_{y}(Y_{1}-Y_{2})}\right],
\nonumber
\end{eqnarray}
with ${\bf R}_{i}=(X_{i},Y_{i})$ the origin of the
cell where the particle $i$ lies. We can verify by direct
inspection that the $W=0$ pair wavefunction $\q^{[A_{2}]}$ vanishes
for $X_{1}=X_{2}$, that is if the particles lie on the same annulus
of the armchair tube.

The effective interaction $W_{\rm eff}$ between the particles of a
$W=0$ pair can be obtained  analytically by the canonical
transformation approach described in Section \ref{section3.2}.
We let $\ve({\bf k})$ be the one-body energy excitation with momentum
${\bf k}$ in the antibonding band and we call ${\cal D}/4$ a
quarter of the empty part of the  FBZ.
The effective Schr\"odinger equation for the pair reads
\begin{equation}
\left[2\varepsilon({\bf k}) +W_{F}+F({\bf k},E)\right]
\,a_{\bf k}+\sum_{{\bf k}'\in {\cal D}/4}\,
W_{\rm eff}\,({\bf k},{\bf k}',E)\,
a_{{\bf k}'} =E a_{\bf k}\;,
\label{cooplike}
\end{equation}
where $W_{F}$ is the first-order self-energy shift (which we found to
be independent of ${\bf k}$) and $F({\bf k},E)$ is the forward scattering
term (which does not contains any direct interaction between the particles
of the pair).
Eq. (\ref{cooplike}) requires a self-consistent calculation of $E$
(since $W_{\rm eff}$ and $F$ are $E$-dependent). We show below that
$E=2\varepsilon_{F}+W_{F}+F_{\rm min}({\bf k}_{F})+\D$, with a positive
binding energy $-\D$ of the $W=0$ pair; here $F_{\rm min}({\bf k}_{F})$
is the minimum value of $F({\bf k},E)$ among the ${\bf k}_{F}$-wavevectors
on the Fermi line.

\begin{figure}[!ht]
\centering
\includegraphics{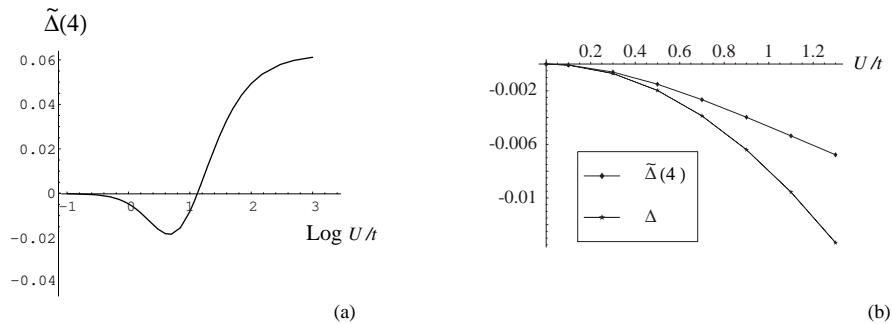}
\caption{(a) Trend of ${\tilde \Delta}(4)$ in units of $t$ versus Log $U/t$ in
the range -1$\div$3 for the (1,1) nanotube. (b) Comparison between
$\D$ and $\tilde{\D}(4)$.}
\label{fig05}
\end{figure}

First, we got a direct verification that pairing actually occurs in
the $(1,1)$ nanotube of length $L=2$ (in units of the lattice spacing) and
periodic boundary conditions. As for the CuO$_{4}$, we compute
the quantity ${\tilde \Delta}(4)=E(4)+E(2)-2\,E(3)$ and we compare it
with $\D$ (obtained from the canonical transformation).
${\tilde \Delta}(4)$ has been computed in a large range of $U/t$ values,
and its trend is shown in Fig. \ref{fig05}(a).
In Fig. \ref{fig05}(b) it is reported the comparison between
$\tilde{\D}(4)$ and $\D$. We observe that the analytical value
$|\Delta|$ is $\sim  2$ times greater than $|\tilde{\D}(4)|$ for
$U/t\simeq 1$, as for the CuO$_{4}$ cluster. However, the
analytical approach predicts the right trend of the
binding energy and it is very reliable in the weak coupling regime.

Next, we consider supercells of $2 \, N \times L= N_{C} $ cells, where $L$
is the length of the $(N,N)$ nanotube in units of the lattice spacing.
We solve the Cooper-like equation (\ref{cooplike}) in a virtually exact way
for $N$ up to 6 and $L$ up to 25, using $U/t=2.5$ (which is of the correct
order of magnitude for graphite\cite{lopez},\cite{hoffman}). The canonical
transformation overestimates $\D$ in this range of $U/t$, but remains
qualitatively correct. The calculations are performed with the Fermi
energy $\varepsilon_{F}$ varying between 0.8 $t$ and 1.1 $t$ (half filling
corresponds to $\varepsilon_{F}=0$). As in the  $(1,1)$ nanotube, the
$W=0$  pairs get bound once dressed by the electron-hole excitations.
However, the binding energy $-\Delta$ decreases monotonically both with
the radius and the length of the tube.

\begin{figure}[!ht]
\centering
\includegraphics{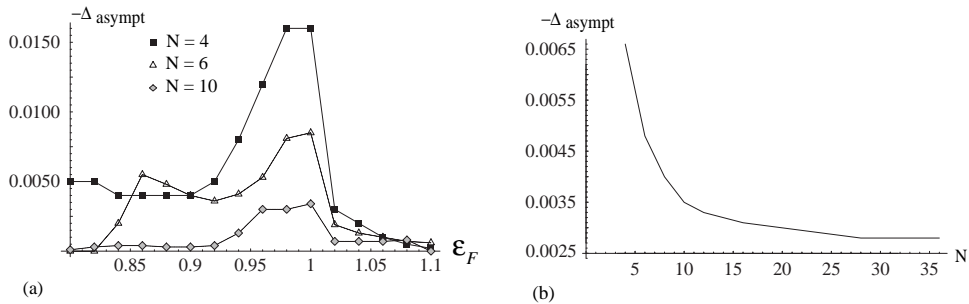}
\caption{(a) Results of the canonical transformation approach with $U/t=2.5$.
$-\Delta_{\rm asympt}$ as a function of the Fermi energy $\varepsilon_{F}$
for $N=4$ (black boxes), $N=6$ (empty triangles) and $N=10$ (grey
diamonds). The Fermi energy varies in the range $0.8\div 1.1$ $t$.
(b) $-\Delta_{\rm asympt}$ as a function of $N$ for $N$
in the range 6$\div$36 with $\varepsilon_{F}=t$ and Average Effective
Interaction $V=1.5\;t$. In both figures $-\Delta_{\rm asympt}$ is in units
of $t$.}
\label{fig06}
\end{figure}

With supercell sizes $N_{C}>300$
numerical calculations become hard and the AEI scheme is used in
order to get reliable extrapolations. The AEI $V_{\rm eff}$ remains fairly
stable around $\approx 1.5\div 2$ $t$ for $N>2$  with increasing  $L$.
Furthermore, $V_{\rm eff}$ is largely independent of the Fermi energy.
The weak dependence of $V_{\rm eff}$ on the lenght $L$ allows for
extrapolating the asymptotic value of the binding energy
$\Delta_{\rm asympt}(N)=\lim_{L \rightarrow \infty}\D(N,L)$, see also
Section \ref{section3.4}. The results are shown in Fig. \ref{fig06}(a)
with $V_{\rm eff}=1.5\;t$. We found that $\Delta_{\rm asympt}$ is strongly
dependent on the filling at fixed $N$; the sharp maximum at the
\textit{optimal doping} $\varepsilon_{F}\approx t$ (which corresponds to a
number of electrons per graphite atom of 1.25) can be understood in terms
of a corresponding peak in the density of states. In the \textit{optimally doped}
case $-\Delta_{\rm asympt}(N)$ decreases monotonically as the radius of the tube
increases, see Fig. \ref{fig06}(b). The decreasing of the binding energy
with $N$ is suggested by recent measurements on nanotubes with diameter
of few Angstrom\cite{tang}. However, in the limit of large $N$,
$\Delta_{\rm asympt}(N)$ remains stable around 0.0028 $t$ and
may be interpreted as the binding energy of the $W=0$ pair in an
\textit{optimally doped} graphite sheet.

We emphasize that our pairing mechanism uses degenerate
electronic states that exist in 2$d$ (or quasi 2$d$) and works
away from half filling. A proper account for the transverse direction
is crucial in order to have a non-Abelian symmetry group and hence
$W=0$ pairs. The $\q^{[A_{2}]}$ pair wavefunction vanishes when the
transverse component $k_{y}=0$. This opens up the interesting
possibility that in nanotubes two distinct superconducting order
parameters appear in the phase diagram, if it turns out that
close to half-filling there is another one which breaks down
the Luttinger liquid.

\subsection{Triangular Lattice in $W=0$ Theory}
\label{section5.2}

We have also considered a symmetric triangular Hubbard lattice, which
may be relevant to the newly discovered\cite{takada} Na$_{x}$CoO$_{2}$
superconductors which are now exciting considerable interest\cite{Lorentz},\cite{chou}.
The 7 atom centered hexagonal cluster with open boundary conditions yields
no pairing for any filling. When opposite sites of this cluster are identified,
one obtains a 4-site cluster which is the smallest one with periodic
boundary conditions. With a hopping integral $t=-1$ we find that
$\tilde{\Delta}(4)$ is negative and shows a similar trend versus $U$ as in the
CuO$_{4}$ case; the pairing energy exceedes $0.1 t$ for $U \sim 4 |t|$.

\section{Superconducting Flux Quantization}
\label{section6}

Bulk superconductors quantize the flux through a hole
in half-integer multiples of the fluxon $\phi_{0}$, because
the quasiparticles that screen the vector potential carry charge
$2e$. In finite systems the signature of superconductivity is  a ground
state energy minimum at $\phi=0$ that is
separated by a barrier from a second minimum at  $\phi=\phi_{0}$/2.  With
increasing the size of the system, the energy (or free energy, at
finite temperature) barrier separating the two minima becomes
macroscopically large, and bulk superconductors can swallow up only
half integer numbers of flux quanta. As emphasized by Canright and
Girvin\cite{cg}, the flux dependence of the ground state energy is
definitely one of the most compelling ways of testing for superconductivity, and
the existence of the two minima separated by a barrier is  a strong
indication of superconducting flux quantization.

\subsection{General Group Theory Aspects of Cu-O Systems}
\label{section6.1}

\begin{figure}[!ht]
\centering
\includegraphics{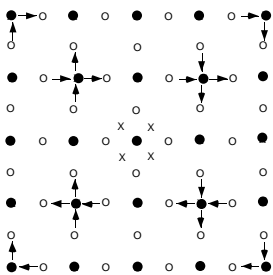}
\caption{Pattern of the vector potential $A$ due to 4 flux
        tubes (crosses) carrying flux $\phi$.
    black dots stand for
    Cu atoms and empty dots stand for O sites.
    The line integral of $A$ along each bond parallel to the arrow is
    $\frac{\phi}{2}$.}
\label{fig07}
\end{figure}

In the present problem, with a repulsive Hubbard model, the mechanism
of attraction is driven by the $C_{4v}$ symmetry, and cannot operate
in an unsymmetric geometry. The flux must be inserted in such a way
that the system is not distorted. In the following we consider Cu-O systems
with $C_{4v}$ symmetry with respect to a central Copper ion, and insert the
magnetic flux in such a way that only 4 central triangular plaquettes
feel a magnetic field (see Figs. \ref{fig07} and \ref{fig08}) and the rest of
the plane only experiences a vector potential.

Consider the pattern of Fig. \ref{fig07}. Here, the Cu sites are
marked by black dots and the Oxygen sites by empty dots; the X's stand
for tubes carrying flux $\phi$ each, symmetrically disposed around the
central Cu.
According with the Peierls prescription we include the magnetic effects
in the Hubbard model by setting
$$
t_{jj'} \to t_{jj'} e^{\frac{2 \pi i}{\phi_{0}} \int_{j}^{j'} {\bf A}
\cdot d {\bf r}} \, ,
$$
where $\phi_{0}=2 \pi /e$ is the flux quantum and $j$ and $j'$ are
the position of the two lattice sites connected by $t_{jj'}$.
Varying $\phi$ by an integer multiple of $\phi_{0}$ corresponds to a gauge
transformation leaving all the physical properties unchanged. The arrows
help to visualise a convenient choice of the gauge at general $\phi$.
Namely, running along an oriented bond in the sense of the arrow,
$\int_{\rightarrow} {\bf A}\cdot d{\bf r}=\frac{\phi}{2}$;
along the other Cu-O bonds, not marked in the figure,
$\int {\bf A}\cdot d{\bf r}=0$. One sees that in this way the
flux through any closed path corresponds to the number of tubes
surrounded by the path. The reflection operations of $C_{4v}$ are
equivalent to $\phi\rightarrow -\phi$, reverse the directions of the
arrows and for a generic $\phi$ the symmetry Group reduces to
$Z_{4}$. However, at $\phi=\frac{\phi_{0}}{2}$ the reversal of the
magnetic field in the tubes corresponds to a jump by $\phi_{0}$, and
this is equivalent to a gauge transformation: this implies that the
symmetry Group gets larger, the new symmetry operations being reflections
supplemented by a gauge transformation. Indeed, the hopping parameter $t_{jj'}$ becomes
$it_{jj'}$ along the arrows, while it remains equal to $t_{jj'}$ along
the unmarked bonds of Fig. \ref{fig07}(a). Any reflection operation simply
changes the signs of all the hoppings along the marked bonds. Now
consider the unitary transformation $S$ which changes the signs of all
the Cu orbitals along both diagonal, except the central Cu. Since $S$
also has the effect of reversing all the arrows, $\sigma \times S$ is a
symmetry, for all reflections $\sigma$ in $C_{4v}$.  Moreover, since
the product of two reflections is a rotation, the Group $\tilde{C}_{4v}$
including the rotations and the reflections multiplied by $S$ is
isomorphic to $C_{4v}$. The $W=0$ pairs appropriate for half a flux quantum must
involve two holes belonging to the degenerate irrep  of $\tilde{C}_{4v}$.
In this way, at  $\phi=\frac{\phi_{0}}{2}$ the full symmetry is
restored, allowing again for  pairing and negative $\Delta$.
If pairing also occurs at $\phi=\frac{\phi_{0}}{2}$, the superconducting
flux quantization arises from a level crossing
between the ground state associated with the paired state at $\phi=0$
and the ground state associated with the paired state at
$\phi=\frac{\phi_{0}}{2}$.

\subsection{Application to the CuO$_{4}$ Case and Numerical Results}
\label{section6.2}

As an illustrative application of the previous symmetry argument, let us
investigate one more time the  CuO$_{4}$ cluster. We expect that this
system is a very good candidate to exhibit superconducting flux quantization,
since it hosts two $W=0$ pairs of different symmetries. As discussed above,
this condition is necessary for the development of a level crossing.
\begin{figure}[!ht]
\centering
\includegraphics{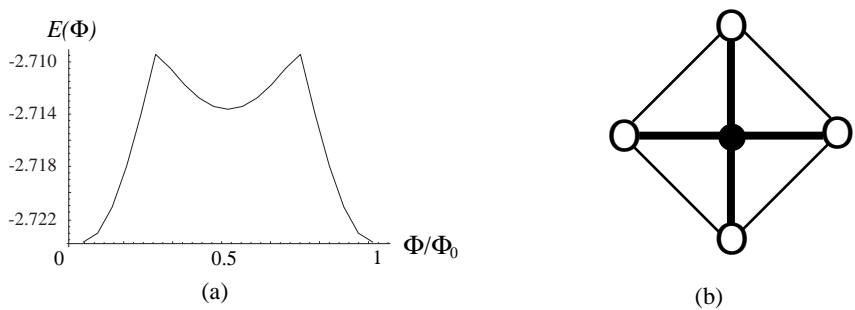}
\caption{(a) Ground state energy of the CuO$_{4}$
with four holes as a function of $\phi/\phi_{0}$. For $\phi/\phi_{0}$
between 0 and 1/4 and between 3/4 and 1 the ground state has $B_{2}$
symmetry; for $\phi/\phi_{0}$ between 1/4 and 3/4 it belong to $A_{1}$. Here
$t=1$eV, $t_{pp}=-0.01$eV, $U=5$eV; $E(4,\phi)$ is in eV. (b)
Topology of the CuO$_{4}$ cluster in presence of $\phi$; the crosses
stand for flux tubes.   }
\label{fig08}
\end{figure}

In agreement with the previous prescription, we have to insert 4 flux
tubes around the central Copper [see Fig. \ref{fig08}(b)] in order to
introduce a closed path around the center, where screening currents can respond.
Every O-O bond collects the Peierls phase
$\frac{2 \pi i }{\phi_{0}}\int {\bf A}\cdot d{\bf r}=
2 \pi i\frac{\phi}{\phi_{0}}$; by symmetry, $t$ is unaffected by the flux.
Thus, the Hamiltonian reads as in Eq. (\ref{sch}) with
$$
t_{pp} \to t_{pp} e^{ 2 \pi i\frac{\phi}{\phi_{0}}} \, .
$$
We observe that a flux of the order of a fluxon in a macroscopic system
would be a small perturbation; in the small cluster, however, the perturbation
is small only if the hopping integral $|t_{pp}|$ is taken  small compared
to $t$. Numerically, the computations were performed with $t_{pp} = -0.01$ eV.
By exact diagonalization, we have found that the ground state
energy $E(\phi)$ of the CuO$_{4}$ with 4 holes, as a function of
$\phi$, has clearly separated minima at zero and half a flux quantum
[see Fig.\ref{fig08}(a)]. Furthermore, $\tilde{\Delta}(4)$ is negative
both at $\phi=0$ and $\phi=\phi_{0}/2$: $\tilde{\Delta}(\phi=0)=-43$meV and
$\tilde{\Delta}(\phi=\phi_{0}/2)=-32$meV.

The physical interpretation of Fig. \ref{fig08}(a) is the following.
When the magnetic flux is inserted into the system, the $W=0$ pair of
$B_{2}$ symmetry
creates a diamagnetic supercurrent that screens the external field. Such a
current flows through the O-O bonds and form closed loops.  As $\phi$ grows
the energy of the system also increases, signaling that the $W=0$ pair is
spending its binding energy to screen the field. At a quarter fluxon
a level crossing occurs, producing a second minimum at $\phi=\phi_{0}/2$.
Here the Hamiltonian is real again and the  $A_{1}$ pair is energetically
favoured. As the flux increases further, the $A_{1}$ pair produce a
new diamagnetic supercurrent until the initial situation is restored
at $\phi=\phi_{0}$. The pairing state at zero flux and half fluxon are orthogonal.
There is a clear analogy with the BCS theory; in that case, the Cooper
wavefunction has $s$ symmetry  and the total magnetic quantum
number of the pair vanishes in the absence of flux, but not at half a
flux quantum.

It is worth to note that if the Hubbard repulsion $U$ is set to zero,
the second minimum at half a fluxon disappears and a trivial
paramagnetic behaviour is observed. This is a further evidence
that the superconducting behaviour of the system is induced by repulsion.

\subsection{Rings of Symmetric Clusters}
\label{section6.3}

We have also studied\cite{bic1},\cite{bic2},\cite{bic3} bound pair hopping and
Superconducting Flux Quantization (SFQ) in systems with CuO$_{4}$ units as
nodes of a graph $\L$. For such
systems the Hamiltonian is $H_{\rm tot} = H_{0}+H_{\t}$ with
$$
H_{0}=\sum_{\a\in \L} [ t \sum_{i\s}( d^{\dag}_{\a \s}p_{\a, i\s}+
p_{\a, i\s}^{\dag}d_{\a\s})+
U(n^{(d)}_{\a \ua} n^{(d)}_{\a \da}+\sum_{i}n^{
(p)}_{\a, i \ua}n^{(p)}_{\a, i\da}
)  ],
$$
while $H_{\t}$ is an intercell hopping Hamiltonian. Here,
$p^{\dag}_{\a, i\s}$ is the  creation operator for a hole
onto the Oxygen $i=1,..,4$ of the $\a$-th cell and so on.
The point symmetry group of $H_{0}$ includes $S_{4}^{|\L|}$,
with $|\L|$  the number of nodes. We shall consider an intercell hopping
which preserves the $S_{4}$  subgroup of $S_{4}^{|\L|}$ in order
to have $W=0$ pair solutions.

\subsubsection{O-O Hopping and SFQ}
\label{section6.3.1}

Consider a
hopping term that allows a hole in the $i$-th Oxygen site of the
$\a$-th unit to move towards the  $i$-th Oxygen site of the
$\b$-th unit with hopping integral
$\t_{\a\b}\equiv|\t_{\a\b}|e^{i\th_{\a\b}}$:
$$
H_{\tau}=\sum_{\a,\b\in\L}\sum_{i\s}
\t_{\a\b}\; p_{\a,i\s}^{\dag}p_{\b,i\s}\; .
$$
For $N=2|\L|$ and $\tau_{\a\b}\equiv 0$, the unique  ground state
consists of 2 holes in each CuO$_{4}$ unit. The intercell hopping produced
by small $|\tau_{\a\b}| \ll |\tilde{\Delta}(4)|$ allows for studying the propagation
of $p$ pairs added to a background $2|\L|$ holes. When $U/t$ is such that
$\tilde{\Delta}(4)<0$, each pair prefers to lie on a single
$\cu$  and for $N=2| \L | + 2p$ the unperturbed ground state is
$2^{p}$$\times$${|\L|}\choose{p}$ times degenerate
(since each CuO$_{4}$ can host two degenerate $W=0$ pairs).
In the low-energy singlet sector, the problem is solved analytically
to second order in $H_{\t}$ and mapped into an effective Hamiltonian
for $p$ hard-core bosons with a complex effective hopping integral
${\cal J}$ that we have calculated analitically and studied as a function
of the ratio $U/t$. For ring-shaped systems, the effective model
is equivalent to the Heisenberg-Ising spin chain
governed by the Hamiltonian
\begin{equation}
H_{\rm HI}=\sum_{\a=1}^{|\L|}{\cal J}
[2\eta\s^{z}_{\a}\s^{z}_{\a+1}+
e^{\frac{4i\p}{|\L|}\frac{\f}{\f_{0}}}\s^{+}_{\a+1}\s^{-}_{\a}+
e^{-\frac{4i\p}{|\L|}\frac{\f}{\f_{0}}}\s^{+}_{\a}\s^{-}_{\a+1}]
\label{heiis}
\end{equation}
where the $\s$'s are Pauli matrices, spin up represents an empty site
and spin down represents a pair. $\eta$ is the so called anisotropy
parameter and in our case $\eta=-1$.
For $\eta=1$, we have the isotropic Heisenberg interaction.
By performing a Jordan-Wigner transformation, the Hamiltonian in
Eq. (\ref{heiis}) can also be mapped into a model of spinless fermions on
the
ring. In the absence of a threading magnetic field ($\f=0$) the problem
was originally studied by Bloch\cite{bloch} and then exactly solved
by Hulthen\cite{hulten} [in the case $\eta=-1$] and
Orbach\cite{orbach} [in the case $\eta\leq -1$] using the Bethe's
hypothesis\cite{bethe}. A systematic analysis in the whole range of
parameters was given by Yang and Yang in a self-contained series of
papers\cite{yang}. Here we just recall that the model has a gapless
phase if $|\eta|\leq 1$, corresponding to the conducting state,
while an insulating phase sets in for $\eta<-1$. As in the 1$d$
Hubbard model, the ``magnetic perturbation'' ($\f\neq 0$) does not
spoil the integrability and the Heisenberg-Ising Hamiltonian remains
exactly solvable by the Bethe-ansatz method. Let us write an
eigenfunction of $H_{\rm HI}$ as
$$
a(\a_{1},...,\a_{p})=\sum_{P}A_{P}\,e^{i\sum_{j}k_{Pj}\a_{j}}
$$
where $P$ is a permutation of the integers $1,\ldots,p$ and $A_{P}$
are $p!$ coefficients. Shastry and Sutherland\cite{shastry} have shown
that the variables $k_{j}$ are given by
\begin{equation}
|\L|k_{j}=2\p I_{j}+4\p\frac{\f}{\f_{0}}-\sum_{l\neq j}\th(k_{j},k_{l})
\label{betheans}
\end{equation}
with a phase shift
$$
\th(k,q)=2\tan^{-1}\left[
\frac{\eta\sin[(k-q)/2]}{\cos[(k+q)/2]-\eta\cos[(k-q)/2]}\right].
$$
From Eqs. (\ref{heiis})-(\ref{betheans}) we readly see that the ground
state energy of the low-energy effective Hamiltonian $H_{\rm HI}$ is
periodic with period $\f_{0}/2$, independent of the number of added
pairs. Thus we conclude that the purely repulsive $\cu$-Hubbard ring
threaded by a magnetic field quantizes the flux in a superconducting
fashion if the number of particles is $2|\L|+2p$ with $0\leq p\leq
|\L|$.

For rings of 2 and  3 $\cu$-sites, we also confirmed the analytic results
by exact diagonalization; for the three-$\cu$ ring  we  observed the SFQ
with total number of holes $2|\L|+2=8$. The switching on of the hopping
$\tau$ between the Oxygen sites breaks the symmetry group $C_{3v}\otimes
S_{4}^{3}$ into $C_{3v} \otimes S_{4}$ for real $\tau$; in a magnetic field
(complex $\tau$), this further breaks into  $C_3 \otimes S_{4}$.
Real $\tau$ lifts the degeneracy between the $k=0$ subspace and the
subspaces $k=1$ and $k=2$ of $C_{3}$ (as usual $k$ is related to the crystal
momentum $p\equiv 2\p\hbar k/3$ in this case), but cannot split
$k=1$ and 2 because they belong to the degenerate irrep of $C_{3v}$;
complex $\tau$ resolves this degeneracy.

\begin{figure}[!ht]
\centering
\includegraphics{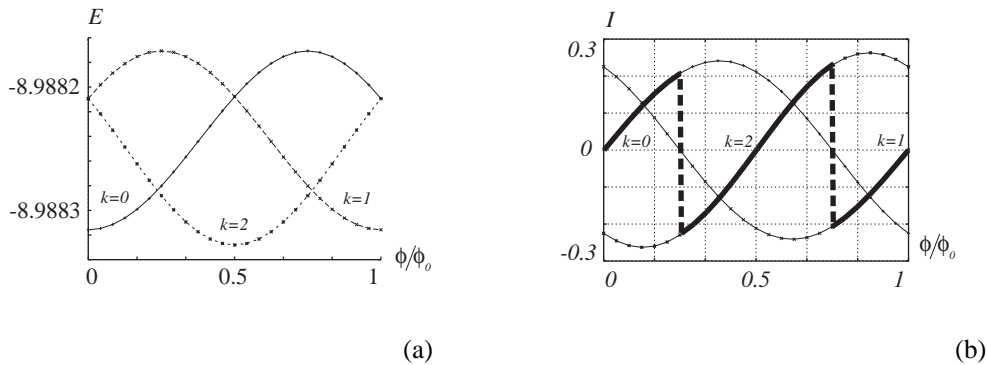}
\caption{(a) Numerical results for the low energy states of the
three-$\cuoq$ ring, as a function of the concatenated magnetic flux.
The energy is in units of $t$. (b) Total current for the three-$\cuoq$
ring, as a function of the magnetic flux. The current is in units of
$e |\tau|/h$. In both figures,
$U=5\,t$, [$\tilde{\Delta}(4) \approx -0.04258 \,t$] $|\tau|=0.001 \,t$.}
\label{fig09}
\end{figure}

The three-$\cuoq$ ring is the smallest ring where we can insert a
magnetic flux $\phi$ by  $\tau=|\tau| e^{i\theta}$,
$\theta=\frac{2 \pi}{3}(\phi/\phi_0)$. The energies of the three
ground-state multiplet components are reported in Fig. \ref{fig09}(a)
for $|\tau|\ll |\tilde{\Delta}(4)|$ and $U=5t$.
At $\f=0$ the ground state belongs to the $k=0$ subspace, while
the first excited levels have $k=1$ and $2$. Their spatial degeneration
is fully lifted: the $k=1$ level increases while the $k=2$ level
decreases up to  $\phi=\phi_{0}/2$.
As $\phi$ increases, the ground state energy grows quadratically in
$\phi$ (diamagnetic behaviour). Near $\phi=\phi_{0}/4$ we
find a level crossing between  $k=0$ and $k=2$, while at
$\phi=\phi_{0}/2$, $k=0$ becomes degenerate with $k=1$ and the ground
state energy is in a new minimum belonging to the $k=2$ subspace:
a sort of ``restoring'' of the $\phi=0$ situation is taking place as
in the BCS theory\cite{lipa}. Indeed, at
$\phi=\phi_{0}/2$ the symmetry group is $\tilde{C}_{3v}\otimes
S_{4}$  where  $\tilde{C}_{3v}$ is isomorphous to
$C_{3v}$ (reflections $\s$ are replaced by $\s g$, where $g$ is a
suitable gauge transformation).

Fulfilling the conditions $\tilde{\Delta}(4)<0$ and $|\tau| \ll |\tilde{\Delta}(4)|$, we have
varied $U$ and $|\tau|$ and found analogous trends for the ground state
energy. Increasing $|\tau|$ with fixed $\tilde{\Delta}(4)$ lowers the central minumum
and depresses the two maxima. On the other hand, if $|\tilde{\Delta}(4)|$ decreases
at fixed $|\tau|$ the central minimum and the side peaks are affected
in a similar way. This is reasonable since the perturbative
parameter is $|\tau|/|\tilde{\Delta}(4)|$.

The current operator $\hat{I} = c \frac{\de H_{0}}{\de \phi}$
yields\cite{kohn2} a gauge invariant average $I$ and in Fig. \ref{fig09}(b)
we display $I$ as a function of $\phi$ in the three ground state sectors
$k=1,\;2$ and 3. The current is proportional to the flux derivative
of the ground-state energy [see Fig. \ref{fig09}(a)] according to
the Hellmann-Feynman theorem and clearly shows a superconducting
behaviour.

The ground state energy in each $k$ sector for the non-interacting
($U=0$)  Cu$_{3}$O$_{12}$ ring shows no pairing in $\cu$ and
indeed the ground state energy is linear in the field at small fields
(normal Zeeman effect). The lowest state is $k=2$ throughout.  The
absence of SFQ is a further evidence of the repulsion-driven pairing mechanism
discussed above. Thus, the dressed $W=0$ pair
screens the vector potential as a particle with an effective charge
$e^{\ast}=2e$ does. At both minima of $E(\f)$ we have
computed for the Cu$_{3}$O$_{12}$ ring $\tilde{\D}(8)\approx -10^{-2} t$.

\subsubsection{Cu-Cu Hopping: infinite effective mass}
\label{section6.3.2}

The three-$\cu$ ring  with $2|\L|+2=8$ holes contains a $W=0$ pair which
in the case $\tilde{\D}(4)<0$ gets bound; an intercell
hopping $\tau_{\rm Cu}$ is now assumed between Cu sites only. This is
perfectly able to carry a one-body current when  the full system is
threaded by the flux, and it does in the absence of interactions. However
the bound pairs have an interesting behaviour.
$|\tau_{\rm Cu}|\ll |\tilde{\D}(4)|$ produces very much smaller
effects on the energy than $\t$ does, but  anyhow
the system behaves as a paramagnet and there is  no flux-induced
splitting of the three $k$ levels.  This happens because   the $W=0$
pair is strictly
localized by the \textit{local} symmetry.  Indeed the $S_{4}$ label of each
$\cu$ unit is
a good quantum number. No SFQ is observed because the screening
of the magnetic field by the bound pair
is forbidden. The superconducting pair behaves as if it had a very large
effective mass.

\section{$W=0$ Pairing and Electron-Phonon Interactions}
\label{section7}

The role of electron-phonon (EP) interactions in determining the
superconducting correlations in the Cu-O planes of cuprates  is a very
controversial issue. Possibly, the pairing mechanism
 has a predominantly electronic origin, but many high-$T_{C}$ compounds exhibit a quite
noticeable doping-dependent isotope effect, suggesting that EP
interactions could be important and should be included in the theory.
In particular there is experimental evidence that the
half breathing Cu-O bond stretching mode at $k=(\pi,0),(0,\pi)$ is significantly coupled
with the doped holes in the superconducting regime and its
contribution may be relevant for the $d_{x^{2}-y^{2}}$
pairing\cite{lanzara}.

Here we further investigate this issue by addressing the question if the
 $W=0$ pairing available in the Hubbard model survives when the
lattice degrees of freedom are switched on.
When lattice effects are introduced in this scenario,  several questions
arise. In the conventional  mechanism, phonons overscreen
the electron repulsion; what happens if electronic screening already
leads to pairing? It is not obvious that  the phonons will reinforce the attraction while
preserving the symmetry.  More generally, some vibrations could be pairing and
others pair-breaking.   To address these problems we use an
extension of the Hubbard model in which bond stretchings dictate the
couplings to the normal modes of the $C_{4v}$-symmetric configuration,
generating a long-range (Fr\"ohlich) EP interaction.

We start from the  Hubbard model with on-site interaction $U$ and
expand the hopping integrals $t_{i,j}({\bf r}_{i},{\bf r}_{j})$ in
powers of the
displacements ${\bf \r}_{i}$ around a $C_{4v}$-symmetric equilibrium configuration
$$
t_{i,j}({\bf r}_{i},{\bf r}_{j}) \simeq t^{0}_{i,j}({\bf r}_{i},{\bf r}_{j})
+ \sum_{\a}\left[ \frac{\partial t_{ij}({\bf r}_{i},{\bf r}_{j}) }
{ \partial r^{\a}_{i}} \right] _{0}   \r^{\a}_{i}+
 \sum_{\a} \left[ \frac{\partial t_{ij}({\bf r}_{i},{\bf r}_{j}) }
{ \partial r^{\a}_{j}} \right] _{0}  \r^{\a}_{j}
\; ,
$$
where  $\a =x,y$.
Below, we  write down the  $\r^{\a}_{i}$  in terms of the normal modes
$q_{\eta \, \nu}$:
$
\r^{\a}_{i}=\sum_{\eta \, \nu} S^{\a}_{\eta \, \nu}(i) \; q_{\eta \, \nu}
$, where $\eta$ is the label of   an irreducible representation ({\em irrep})  of the
symmetry group of the undistorted system and $\n$ is a phonon branch.

Thus,  treating  the  Cu  atoms as fixed, for
simplicity, one can justify an electron-lattice Hamiltonian:
$$
H_{el-latt} =  H_{0} + V_{\rm tot} \, .
$$
Here $ H_{0} = H_{0}^{n}+H_{0}^{e}$ accounts for the kinetic part of
the electron-phonon system and it is given by
$$
  H_{0} =\sum_{\eta} \hbar \omega_{\eta,\n}
  b^{\dagger}_{\eta,\n}
  b_{\eta,\n}+
  \sum_{i,j\s} t^{0}_{i,j}({\bf r}_{i},{\bf r}_{j})\; c^{\dag}_{i
  \s}c_{j \s};
 $$
the $\omega$'s are normal mode frequencies. The interacting
part $ V_{\rm tot} = V+W$ contains the Hubbard repulsion $W$ and the
electron-phonon interaction $V$. The latter can be written in terms
of the parameters $\xi_{\eta,\nu}=\lambda_{\eta\,\nu}\sqrt{\frac{\hbar}
{2M\omega_{\eta,\n}}}$, where the $\l$'s are numbers of order unity
that modulate the EP coupling strength, while $M$ is the Oxygen mass.
Then,
$$
V=\sum_{\eta,\n}\xi_{\eta,\nu} (
b^{\dagger}_{\eta,\n}+ b_{\eta,\n}) H_{\eta, \n},
$$
and the $H_{\eta,\n}$ operators  are given by
$$
 H_{\eta,\n}=\sum_{i,j} \sum_{\a, \sigma} \left\{ S^{\a}_{\eta \, \nu}(i)
 \left[ \frac{\partial t_{ij}({\bf r}_{i},{\bf r}_{j}) }
{ \partial r^{\a}_{i}} \right] _{0}  \right.
\left.+ S^{\a}_{\eta \, \nu}(j)
\left[ \frac{\partial t_{ij}({\bf r}_{i},{\bf r}_{j}) }
{ \partial r^{\a}_{j}} \right] _{0}    \right\} ( c^{\dag}_{i,
\s}c_{j ,\s}+h.c.) \, .
$$
 This is
physically more detailed than the  Hubbard-Holstein model,
where electrons are coupled to a local Einstein phonon and the
superconducting phase has been investigated in detail\cite{pao},\cite{alexandrov}.
Indeed, in the present context the Hubbard-Holstein model is not fully satisfactory because it
 restricts to on-site EP couplings, which would be impaired by a
strong Hubbard repulsion. On the other hand the use of a Fr\"ohlich-like
phonons was suggested\cite{alexandrov2},\cite{alexandrov3},\cite{bonca2}
for modeling the Cu-O  planes in the strong
EP coupling regime: a long-range EP coupling removes the problem of  polaron
self-trapping, otherwise present in the case of the Holstein interaction,
where an unphysically large polaron (and bipolaron) mass occurs.

By   generalizing to  $H_{el-latt}$
the canonical transformation  proposed for the electronic part
$H=
H_{0}^{e}+W$, one can derive\cite{pcfon1},\cite{pcfon2}  an
effective interaction between the particles in the pair.
We obtained a new  Cooper-like equation  $H_{\rm pair}|\varphi \rangle  = E |\varphi \rangle  $
with an effective two-body Hamiltonian, acting upon the dressed $W=0$
pair $|\varphi \rangle  $.
As in Section 3 the pairing criterion involves the properly renormalized  Fermi
energy $E_{F}^{(R)}$:
if the lowest energy eigenvalue $E$ is such that
$E=E_{F}^{(R)}+\D$ with negative
$\D$, the dressed $W=0$ pair gets bound in the many-body
interacting problem and the system undergoes a  Cooper instability.
We observed that this extended approach is very accurate at weak coupling and is
qualitatively predictive also in the intermediate coupling regime.

As an illustrative application  we focussed again  on CuO$_{4}$.
This cluster  represents a  good test of the interplay
between electronic and phononic pairing
mechanisms since we can compare exact diagonalization
results with the analytic approximations of the canonical
transformation. CuO$_{4}$ allows   only the coupling to phonons at the centre
or at  the edge of
the Brillouin Zone;  however, phonons near the edge are precisely the  most
relevant ones\cite{lanzara}.
We recall (see Section \ref{section2.2}) that the pure Hubbard CuO$_{4}$ cluster with O-O
hopping    $t_{pp}=0$ yields ${\tilde \Delta} (4) <0$,
and a degenerate   $W=0$ bound pair  with $A_{1}$
and  $B_{2}$ components. A standard Jahn-Teller calculation in which the
degenerate ground state wave functions interact with the vibrations predicts
distortions that already at moderate EP coupling destroy the
pairing\cite{pcfon2}.
We also explored the scenario beyond this approximation.
The analytical solution of the Cooper-like equation shows that the
electronic pairing is enhanced
by the phonon contribution. In particular the  binding energy
of the pair in the $B_{2}$ symmetry channel increases more rapidly
than in the $A_{1}$ channel.

In order to go beyond the weak-coupling regime
we have exactly diagonalized  the Hamiltonian in a truncated Hilbert space.
We obtained several indications. First, the catastrophic distortions
predicted by the  Jahn-Teller
approximation are not borne out by  more complete
approaches involving  a broader spectrum of electronic states.
Pairing prevails also at strong  coupling
in part of the parameter space, in the symmetry channels where $W=0$
pairs occur. The correct trend is predicted by the canonical
transformation approach, which also explains the pairing or
pair-breaking character of the individual normal modes. In particular it is found that
the half-breathing modes give a  synergic contribution to the purely
electronic pairing; since   they are believed to be   mainly involved in
optimally doped cuprates, our findings suggest  a joint
mechanism, with the Hubbard model that captures a crucial part of the physics.

\section{Conclusions and Outlook}
\label{section8}

We have presented the following evidence that the $W=0$ pairs are
the quasi-particles that, once dressed, play the r\^{o}le of
Cooper pairs: 1) as two-body states they do not feel the large
on-site repulsion, that would come in first-order perturbation in
any theory of pairing with any other kind of pairs. 2) The
indirect interaction with the background particles gives
attraction, and bound states with physically appealing binding
energies. 3) The same results are also borne out by exact
diagonalization in finite clusters, if and only if they have the
correct symmetry and filling to give raise to $W=0$ pairs in
partially filled shells. 4) Both in clusters and in the plane, the
superconducting flux quantization results from the symmetry
properties of $W=0$ pairs.

The setup of our theory of  the effective interaction $W_{\rm
eff}$ between two holes (or electrons) is quite general; although
we developed it in detail for Hubbard Models, it is well suited to
be extended in order to  include other ingredients, like phonons.
In principle we can obtain $W_{\rm eff}$ by our canonical
transformation, including systematically  any kind of virtual
intermediate states.  The closed-form analytic expression of
$W_{\rm eff}$ we obtained includes  4-body virtual states. This
describes repeated exchange of an electron-hole pair to all
orders. One can envisage the pairing mechanism by spin-flip
exchange diagrams that are enhanced by the $C_{4v}$ symmetry. The
argument does not depend in any way on perturbation  theory, and
the equations retain their form, with renormalized parameters, at
all orders. We find that in the three-band Hubbard Model,
$^{1}A_{2}$ pairs are more tightly bound close to half filling,
but $^{1}B_{2}$ pairs are favored when the filling increases.
Here, like in the BCS theory, the superconducting flux
quantization relies on the existence of bound pairs of different
symmetries.  We recall, however, that these symmetry labels are
not absolute, but depend on the choice of a gauge convention. We
get attraction and pairing at all fillings we have considered
(above half filling), but the binding energy of the $^{1}A_{2}$
pairs drops by orders of magnitude as the filling increases. The
three-band Hubbard Model might be too idealized to allow a
detailed quantitative comparison with experiments; however some
qualitative answers are very clear.

Using the Lieb's theorem\cite{lieb} and a symmetry analysis based
on the $W=0$ theorem, one can fix the way the interactions remove
the high degeneracy of the Hubbard Model at half filling. This can
be used to study pairing in the doped system by the above
canonical transformation.    Exact  numerical data on the $4\times
4$ square lattice are available\cite{fop}, and this allows to test
the $W=0$ pairing mechanism within the one-band Hubbard Model,
improving over previous  weak coupling
analyses\cite{fridman}\cite{galan}. For a full application of the
W=0 theorem, we had to  include nontrivial symmetries which are no
isometries; this enabled us to  classify the $W=0$ channels and
calculate  the binding energies analytically.  Once the analysis
is so complete, the criteria for pairing are unambiguous. Moreover
the success of the weak coupling approach has been noted by other
authors\cite{kondo}, and the reason is fully clarified by the
$W=0$ theory.

Despite some evident and important differences that we pointed out
above, our mechanism can be thought of broadly speaking as a
lattice realization of the pioneering one proposed long
ago\cite{kohn} by Kohn and Luttinger. The repulsion is avoided by
the  $W=0$ configuration mixing without using parallel spins and
high angular momenta. The basic source of attraction, however,  is
essentially the same, since in second-order, in the singlet
channel, the spin-flip diagram is the only one that survives. We
found pairing in  the singlet channel in a variety of models
including carbon nanotubes\cite{PRB2002}.  This real-space
approach is suitable for realizing thought experiments, like those
involving the SFQ (Section \ref{section6}).

On the other hand, the above results also prove that important
ingredients are still missing and must be included. The $4\times 4$
model shows evidence of bound pairs of non-vanishing momentum, in
degenerate representations. This opens up the possibility of charge
inhomogeneities and Jahn-Teller distorsions.
However modeling vibration effects in a CuO$_{4}$ cluster we find
that the outcome depends on which phonon mode is most strongly
coupled. The bound pairs can be flexible enough to prevent distortions
and actually gain binding energy in the presence of $A$ and $B$
phonons, whilst $E$ vibrations are definitely pair breaking.

\begin{center}
\bigskip

\end{center}


\bigskip
\section*{References}

\begin{thebibliography}{100}

\bibitem{bedmu}
J. G. Bednorz and K. A. M\"{u}ller, Z. Phys. {\bf B 64}, 189 (1986).
\bibitem{fullerene}
J. H. Schšn, Ch. Kloc, R. C. Haddon, and B. Batlogg, Science {\bf 288}, 656 (2000).
\bibitem{tang}
Z. K. Tang, Lingyum Zhang, N. Wang, X. X. Zhang, G. H. Wen, G. D. Lee, J. N. Wang,
C. T. Chan and Ping Sheng, Science {\bf 292}, 2462 (2001).
\bibitem{ruthe}
Y. Maeno et al., Nature (London) {\bf 372}, 532 (1994).
\bibitem{takada}
K. Takada, H. Sakurai, E. Takayama-Muromachi, F. Izumi, R. A. Dilanian,
T. Sasaki,  Nature {\bf 422}, 53 (2003).
\bibitem{hubbard}
J.~Hubbard,
Proc. Roy. Soc. A {\bf 276}, 238 (1963);
Proc. Roy. Soc. A {\bf 277}, 237 (1964);
Proc. Roy. Soc. A {\bf 281}, 401 (1964).
\bibitem{bickers2}
N. E. Bickers, D. J. Scalapino and S. R. White, Int. J. Mod. Phys. {\bf 1}, 687 (1987).
\bibitem{hotta}
Yoichi Yanase, Takanobu Jujo, Takuji Nomura, Hiroaki
Ikeda, Takashi Hotta and Kosaku Yamada, Physics Reports {\bf 387}, 1-149 (2003).
\bibitem{dagrev}
For a review on the t-J Model see, \textit{e.g.}, E. Dagotto, Rev. Mod. Phys. {\bf 66}, 763 (1994).
\bibitem{sush}
O. P. Sushkov, Phys. Rev. {\bf B 54}, 9988 (1996) and reference therein;
V. V. Flambaum, M. Y. Kuchiev and O. P. Sushkov, Physica {\bf C 235-240}, 2218 (1994).
\bibitem{gutz}
M. C. Gutzwiller, Phys. Rev. Lett. {\bf 10}, 159 (1963).
\bibitem{kohn}
W. Kohn and J. M. Luttinger, Phys. Rev. Lett. {\bf 15}, 524 (1965).
\bibitem{shankar}
R. Shankar, Rev. Mod. Phys. {\bf 66}, 129 (1994).
\bibitem{chubukov}
A. V. Chubukov, Phys. Rev. B {\bf 48}, 1097 (1993).
\bibitem{bickers}
N. E. Bickers, D. J. Scalapino and S. R. White, Phys. Rev. Lett.
{\bf 62}, 961 (1989); P. Monthoux, A. Balatsky and D. Pines, Phys.
Rev. Lett. {\bf 67}, 3448 (1991); N. E. Bickers and S. R. White,
Phys. Rev. {\bf B 43}, 8044 (1991).
\bibitem{zanchi}
D. Zanchi and H. J. Schulz, Phys. Rev. {\bf B 54}, 9509 (1996);
D. Zanchi and H. J. Schulz, Phys. Rev. {\bf B 63}, 13609 (2000);
\bibitem{metzner}
C. J. Halboth and W. Metzner, Phys. Rev. Lett. {\bf 61}, 7364 (2000).
\bibitem{honerkamp}
C. Honerkamp and M. Salmhofer, Phys. Rev. Lett. {\bf 64}, 184516 (2001).
\bibitem{anderson3}
P.~W. Anderson, G. Baskaran, Z. Zou and T. Hsu, Phys. Rev. Lett. {\bf 58}, 2790 (1987).
\bibitem{anderson4}
P.~W. Anderson and Z. Zou, Phys. Rev. {\bf 60}, 132 (1988).
\bibitem{susu}
G. Su and M. Suzuki, Phys. Rev. {\bf B 58}, 117  (1998).
\bibitem{liebwu}
E. H. Lieb and F. Y. Wu, Phys. Rev. Lett. {\bf 20}, 1445 (1968).
\bibitem{1dl}
The Hubbard chain is classified as a  Mott insulator with a spin gap at half-filling
and as a Luttinger liquid away from it.
\bibitem{2dl}
Among them we mention the theorems by Lieb, Ref.\cite{lieb}, on the ground state spin
degeneracy at half filling, the ferromagnetic ground state solutions
devised by Nagaoka, Ref.\cite{nagaoka}, and by Mielke, Ref.\cite{mielke},
and Tasaki, Ref.\cite{tasaki}, and the extensions of Mermin-Wagner theorem on the
absence of long-range order at finite temperature, provided by Su and
Suzuki, Ref.\cite{susu}.
\bibitem{lieb}
E.~H. Lieb, Phys. Rev. Lett. {\bf 62}, 1201 (1989).
\bibitem{nagaoka}
Y. Nagaoka, Phys. Rev. {\bf 147}, 392 (1966).
\bibitem{mielke}
A. Mielke, J. Phys. A {\bf 24}, L73, (1991);
A. Mielke, J. Phys. A {\bf 25}, 4335 (1992);
A. Mielke, Phys. Lett. A {\bf 174}, 443 (1992).
\bibitem{tasaki}
H. Tasaki, Phys. Rev. Lett. {\bf 69}, 1608 (1992);
H. Tasaki, Phys. Rev. B {\bf 40}, 9192 (1989).
\bibitem{richardson}
R. W. Richardson, Phys. Rev. {\bf 141}, 949 (1966).
\bibitem{balseiro}
C. A. Balseiro, A. G. Rojo, E. R. Gagliano and B. Alascio,
Phys. Rev. B {\bf 38}, 9315 (1988).
\bibitem{gubernatis}
S. Zhang, J. Carlson and J. E. Gubernatis,
Phys. Rev. Lett. {\bf 84}, 2550 (2000).
\bibitem{fop}
G. Fano, F. Ortolani and A. Parola, Phys. Rev. B {\bf 42}, 6877 (1990);
A. Parola, S. Sorella, M. Parrinello and E. Tosatti, Phys. Rev. B {\bf 43}, 6190 (1991);
G. Fano, F. Ortolani and A. Parola, Phys. Rev. B {\bf 46}, 1048 (1992);
G. Fano, F. Ortolani and F. Semmeria, Int. J. Mod. Phys. B {\bf 3}, 1845 (1990);
\bibitem{size}
The $4\times 4$ model is the largest square Hubbard system where exact
diagonalizations have been performed; the Hilbert space contains indeed
$166 \times 10^{6}$ configurations!
\bibitem{cb1}
M. Cini and A. Balzarotti, Il Nuovo Cimento D {\bf 18}, 89 (1996).
\bibitem{scalah}
J. E. Hirsch, S. Tang, E. Loh and D. J. Scalapino, Phys. Rev. Lett. {\bf 60}, 1668 (1988).
\bibitem{scalah2}
J. E. Hirsch, S. Tang, E. Loh and D. J. Scalapino, Phys. Rev. B {\bf 39}, 243 (1989).
\bibitem{ogata}
M. Ogata and H. Shiba, J. Phys. Soc. Jpn. {\bf 57}, 3074 (1988).
\bibitem{martin}
R. L. Martin, Phys. Rev. B {\bf 14}, R9647 (1996).
\bibitem{emery2}
V. J. Emery and G. Reiter, Phys. Rev. B {\bf 38}, 4547 (1988).
\bibitem{fulde}
P. Fulde, {\tt Electron Correlation in Molecules and
Solids}, Springer-Verlag series in Solid State Sciences 100 (1991).
\bibitem{hyber}
M. S. Hybertsen, M. Schl\"{u}ter and N. Christensen, Phys. Rev. B {\bf 39}, 9028 (1988);
M. Schl\"{u}ter, in {\tt Superconductivity and Applications}, edites by H. S. Kwok
{\it et al.} (Plenum Press, New York, N. Y.) p. 1 (1990).
\bibitem{zr}
F. C. Zhang and T. M. Rice, Phys. Rev. B {\bf 37}, 3759 (1988).
\bibitem{cb34}
M. Cini and A. Balzarotti, Solid State Comm. {\bf 101}, 671 (1997);
M. Cini, A. Balzarotti, J. Tinka Gammel and A. R. Bishop, Il Nuovo
Cimento D {\bf 19}, 1329 (1997).
\bibitem{EPJB2000}
M. Cini, A. Balzarotti and G. Stefanucci, Eur. Phys. J. B {\bf 14}, 269 (2000).
\bibitem{cb5}
M. Cini and A. Balzarotti, Phys. Rev. B {\bf 56}, 14711 (1997).
\bibitem{ijmp00}
M. Cini, A. Balzarotti, R. Brunetti, M. Gimelli and G. Stefanucci,
Int. J. Mod. Phys. B {\bf 14}, 2994, (2000).
\bibitem{metzner2}
W. Metzner and D. Vollhardt, Phys. Rev. B {\bf 39}, 4462 (1989).
\bibitem{fridman}
B. Friedman, Europhys, Lett. {\bf 14}, 495 (1991).
\bibitem{galan}
J. Galan and J. A. Verges, Phys. Rev. B {\bf 44}, 10093 (1991).
\bibitem{optimal}
Consider for instance that the largest irrep of the space
group has dimension 8, while the half-filling shell is $2N-2$
times degenerate, where $N \times N$ is the total number of unit
cells in the system.
\bibitem{EPJB1999}
M. Cini, G. Stefanucci and A. Balzarotti, Eur. Phys. J. B {\bf 10}, 293 (1999).
\bibitem{SSC1999}
M. Cini, G. Stefanucci and A. Balzarotti, Solid State Commun. {\bf 109}, 229 (1999).
\bibitem{obhm}
Similarly, we have shown that the One Band Hubbard model admits $W=0$ pairs
belonging to the $A_{2}$, $B_{1}$ and $B_{2}$ irreps.
\bibitem{cbs1}
M. Cini, A. Balzarotti and G. Stefanucci, {\tt cond-mat/9811116}.
\bibitem{e-hsymmetry}
For the sake of clarity we are assuming here that the particles we are considering
are holes, as in cuprates; however the canonical transformation is
not limited to this case and it is applicable both to electrons and holes.
\bibitem{shen1}
Z. X. Shen, W. E. Spicer, D. M. King, D. S. Dessau and B. O. Wells,
Science {\bf 267}, 343 (1995) and references therein.
\bibitem{lynch}
D. W. Lynch and C. G. Olson, {\tt Photoemission Studies of High-Temperature Superconductors},
Cambridge Univ. Press. Cambridge (U.K.) 1999.
\bibitem{sc}
G. Stefanucci and M. Cini, Phys. Rev. B {\bf 66}, 115108 (2002).
\bibitem{sqt}
S. Q. Shen, Z. M. Qiu and G. S. Tian, Phys. Rev. Lett. {\bf 72}, 1280 (1994).
\bibitem{SSC2001}
M. Cini and G. Stefanucci, Solid State Comm. {\bf 117}, 451 (2001).
\bibitem{jop2001}
M. Cini and G. Stefanucci, J. Phys. C {\bf 13}, 1279  (2001).
\bibitem{moreodag}
A. Moreo and E. Dagotto, Phys. Rev. B {\bf 41}, 9488 (1990).
\bibitem{bonca}
J. Bonca, P. Prelovsek and I. Sega, Phys. Rev. B {\bf 39}, 7074 (1989).
\bibitem{poilb}
Y. Hasegawa and D. Poilblanc, Phys. Rev. B {\bf 40}, 9035  (1989).
\bibitem{EPJB2001}
M. Cini, E. Perfetto and G. Stefanucci, Eur. Phys. J. B {\bf 20}, 91 (2001).
\bibitem{belash}
J. T. Belash, O. V. Zharikov and A. V. Pal'nichenko, Solid State Comm. {\bf 63}, 153 (1987);
J. T. Belash, A. D. Bronnikov, O. V. Zharikov and A. V. Pal'nichenko, Solid State Comm. {\bf 64}, 1445 (1987);
J. T. Belash, A. D. Bronnikov, O. V. Zharikov and A. V. Pal'nichenko, Solid State Comm. {\bf 69}, 921 (1989).
\bibitem{bockrath2}
M. Bockrath, J. Hone, H. Zettl, P. L. McEuen, A. G. Rinzler and R.E. Smalley,
Phys. Rev. B {\bf 61}, R10606 (2000).
\bibitem{gao}
B. Gao, A. Kleinhammes, X. P. Tang, C. Bower, L. Fleming, Y. Wu and O. Zhou,
Chem. Phys. Lett. {\bf 307}, 153 (1999);
J. Zhao, A. Buldum, J. Han and J. Ping Lu, Phys. Rev. Lett. {\bf 85}, 1706 (2000).
\bibitem{zhao}
J. Zhao et al., Phys. Rev. Lett. {\bf 85}, 1706 (2000).
\bibitem{balents}
L. Balents and M. P. A. Fisher, Phys. Rev. B {\bf 55}, R 11973 (1997).
\bibitem{sol}
Jen\"o S\'olyom, in \textit{Proceedigs of the School and Workshop
on Nanotubes and Nanostructures 2000}, Santa Margherita di Pula (Ca), Italy.
\bibitem{PRB2002}
E. Perfetto, G. Stefanucci, and M. Cini, Phys. Rev. B {\bf 66}, 165434 (2002);
E. Perfetto , G. Stefanucci , M. Cini, Eur. Phys. J. B {\bf 30}, 139 (2001).
\bibitem{lopez}
M. P. Lopez Sancho, M. C. Munoz and L. Chico, Phys. Rev. B {\bf 63}, 165419 (2001).
\bibitem{hoffman}
A. L. Tchougreef and R. Hoffman, J. Phys. Chem. {\bf 96}, 8993 (1992).
\bibitem{Lorentz}
B. Lorenz, J. Cmaidalka, R. L. Meng, C. W. Chu, Phys. Rev. B {\bf 68}, 132504 (2003).
\bibitem{chou}
F. C. Chou, J. H. Cho, Patrick A. Lee, E. T. Abel, K. Matan, Young S. Lee, Phys. Rev. Lett.
{\bf 92}, 157004 (2004).
\bibitem{cg}
G. S. Canright and S. M. Girvin, Int. J. Mod. Phys. B {\bf 3}, 1943 (1989).
\bibitem{bic1}
Michele Cini, Gianluca Stefanucci, Enrico Perfetto and Agnese Callegari,
J. Phys. C {\bf 14} L709 (2002).
\bibitem{bic2}
A. Callegari, M. Cini, E. Perfetto and G. Stefanucci, Eur. Phys. J. B {\bf 34}, 455 (2003).
\bibitem{bic3}
A. Callegari, M. Cini, E. Perfetto and G.Stefanucci, Phys. Rev. B {\bf 68}, 153103 (2003).
\bibitem{bloch}
F. Bloch, Z. Physik {\bf 61}, 206 (1930); F. Bloch, Z. Physik {\bf 74}, 295 (1932).
\bibitem{hulten}
L. Hulthen, Arkiv. Mat. Astron. Fysik {\bf 26 A}, No. 11 (1938).
\bibitem{orbach}
R. Orbach, Phys. Rev. {\bf 112}, 309 (1958).
\bibitem{bethe}
H. A. Bethe, Z. Physik {\bf 71}, 205 (1931).
\bibitem{yang}
C. N. Yang and C. P Yang, Phys. Rev. {\bf 147}, 303 (1966);
C. N. Yang and C. P Yang, Phys. Rev. {\bf 150}, 321 (1966);
C. N. Yang and C. P Yang, Phys. Rev. {\bf 150}, 327 (1966);
C. N. Yang and C. P Yang, Phys. Rev. {\bf 151}, 258 (1966).
\bibitem{shastry}
B. S. Shastry and B. Sutherland, Phys. Rev. Lett. {\bf 65}, 243 (1990);
B. Sutherland and B. S. Shastry, Phys. Rev. Lett. {\bf 65}, 1833 (1990).
\bibitem{lipa}
W. A. Little and R. D. Parks, Phys. Rev. Lett. {\bf 9}, 9 (1962).
\bibitem{kohn2}
W. Kohn, Phys. Rev. {\bf  133}, A171 (1964).
\bibitem{lanzara}
Z. X. Shen, A. Lanzara, S. Ishihara, and N. Nagaosa, Phil. Magazine B 82, 1349-1368 (2002).
\bibitem{pao}
C. H. Pao and H. B. Schuttler, Phys. Rev. B {\bf 57}, 5051 (1998);
Phys. Rev. B {\bf 60}, 1283 (1999).
\bibitem{alexandrov}
A. S. Alexandrov, Phys. Rev. B {\bf 46}, 2838 (1992);
G. Wellein, H. Roder and H. Fehske, Phys. Rev. B {\bf 53}, 9666 (1996);
J. Bonca, T. Katrasnik and S.A. Trugman, Phys. Rev. Lett. {\bf 84}, 3153 (2000).
\bibitem{alexandrov2}
A. S. Alexandrov and P. E. Kornilovitch, J. Phys. C {\bf 14},5337 (2002).
\bibitem{alexandrov3}
A. S. Alexandrov, Int. J. Mod.  Phys. B {\bf 17}, 3315 (2003).
\bibitem{bonca2}
J. Bonca and S. A. Trugman, Phys. Rev. B {\bf 64}, 094507 (2001).
\bibitem{pcfon1}
E. Perfetto and M. Cini, Phys. Rev. B {\bf 69}, 92508 (2004).
\bibitem{pcfon2}
E. Perfetto and M. Cini, J. Phys. C {\bf 16} 4845 (2004).
\bibitem{kondo}
Jun Kondo, J. Phys. Soc. Japan {\bf 70}, 808 (2001).



\end{thebibliography}
\end{document}